\documentclass[12pt]{iopart}

\usepackage{iopams}

\usepackage{amssymb}
\usepackage{framed}

\usepackage[dvips]{graphicx}

\def\II{\mathbb I}

\def\Pr{{\rm Pr}}

\newtheorem{Lmm}{Lemma}
\newtheorem{Thm}{Theorem}
\newtheorem{Dfn}{Definition}
\newtheorem{Crl}{Corollary}

\newcommand{\sq}{\hspace{\fill}$\blacksquare$}

\begin{document}

\title[Multi-partite squash operation and DIQKD]{Multi-partite squash operation and its application to device-independent quantum key distribution}

\author{Toyohiro Tsurumaru$^1$ and Tsubasa Ichikawa$^2$\footnote{Present address: Advanced Data Solutions,
Corporate Marketing Department, Teikoku Data Bank, LTD., 2-5-20, Minami-Aoyama, Minato-ku, Tokyo, 107-8680.}}

\address{
$^1$ Mitsubishi Electric Corporation,
Information Technology R\&D Center, 5-1-1 Ofuna, Kamakura-shi, Kanagawa, 247-8501, Japan}
\address{
$^2$ Department of Physics, Gakushuin University, Mejiro, Toshima-ku, Tokyo, 171-8588, Japan}
\vspace{10pt}
\begin{indented}
\item[]September 2015
\end{indented}

\begin{abstract}
The squash operation, or the squashing model, is a useful mathematical tool for proving the security of quantum key distribution systems using practical (i.e., non-ideal) detectors.
At the present, however, this method can only be applied to a limited class of detectors, such as the threshold detector of the Bennett-Brassard 1984 type.
In this paper we generalize this method to include multi-partite measurements, such that it can be applied to a wider class of detectors.
We demonstrate the effectiveness of this generalization by applying it to the device-independent security proof of the Ekert 1991 protocol, and by improving the associated key generation rate.
For proving this result we use two physical assumptions, namely, that quantum mechanics is valid, and that Alice's and Bob's detectors are memoryless.
\end{abstract}

%
%
%
%
%

\section{Introduction}

Quantum key distribution (QKD) \cite{BB84} is a technique for distributing information-theoretically secure secret keys between two parties connected by a quantum channel.
Beginning from the Bennett-Brassard 1984 (BB84) \cite{BB84}, and the Ekert 1991 protocols \cite{E91}, there is now a variety of  protocols proposed, e.g., \cite{B92,IWY02,GG02,HYAKN03,SYM14}.
Several different approaches have been advanced for proving the security of QKD systems using the ideal qubit detectors \cite{SP00,KGR05, Koashi}.

The squash operation, or the squashing model, is a useful mathematical tool for proving the security of QKD systems using practical (i.e., non-ideal) detectors \cite{TT08,BML08}.
Once its existence is proved for a given practical detector, one can incorporate it into a conventional type of security proof where receivers have ideal qubit detectors, and automatically obtains a new proof that remains valid even if the practical detectors are used.
The squash operation literally {\it squashes} an incoming state to a qubit, and also has a property that, when followed by qubit measurements, it acts exactly the same way as the practical detector.
In security proofs, there is no loss of generality in supposing that the squash operation is conducted by the attacker, and as the result of that, the security of a protocol using practical detectors is reduced to that using ideal qubit detectors.

A type of squash operation was first assumed in the security proof by Gottesman et al. \cite{GLLP02}, however, its existence was only conjectured, no proof was given.
The first proof was given by one of the present authors and Tamaki \cite{TT08}, for the case of the threshold detector of the BB84 type measurement.
This result was also verified independently by Beaudry, Moroder, and L\"utkenhaus \cite{BML08}.
There were also efforts toward constructing squash operations for a wider class of practical detectors.
For example, Beaudry et al. gave an explicit condition for the existence of a squash operation, and used it to show positive and negative results on the six-state protocol with threshold detectors \cite{BML08}.
Later their techniques were refined further and applied to other types of measurement devices \cite{GBNAML14}.
In Ref. \cite{T10}, one of the present authors discussed whether symmetries of a given detector can imply the existence of the squash operation corresponding to it, and also showed that the above result on the BB84 type measurement is valid even for multi-mode cases.
In addition to these uses in quantum cryptography, Moroder et al. applied the squash operation to entanglement verification with realistic measurement devices \cite{MGBPL10}.

Despite all these efforts, however, the method of the squash operation is still applicable only to a limited class of measurement devices.
In fact, even if we restrict ourselves to qubit measurements of the BB84 type, we can easily construct counterexamples to its existence (see Lemma \ref{lmm:no-go_theorem} in Section \ref{sec:multi-partite_squash_operations}).

In this paper, we demonstrate that the situation changes drastically by considering a generalized case where multi-partite measurements are involved.
That is, while all previous studies on the squash operation were concerned only with detectors used by a single player, we here consider a generalization including global measurements performed jointly by two players or more, such as the Clauser-Horne-Shimony-Holt (CHSH) measurement \cite{CHSH69}, used e.g. in the E91 protocol.
This approach allows us to relax mathematical conditions required for the existence of the squash operation, such that they can be fulfilled for a wider class of detectors.
Perhaps this is most easily illustrated by considering the CHSH measurement as an example.
If one regards the CHSH measurement as a mixture of local  $x,z$-basis measurements performed by Alice and Bob, there are two  basis for each player, which together yield four conditions that the squash operation has to satisfy.
On the contrary, if one regards the same measurement as one global measurement, there is no basis choice, and thus only one condition is required for the existence of the squash operation.

As an evidence of the effectiveness of this generalization, we apply it to the device-independent security analysis of the E91 protocol, and improve the key generation rate known so far:
The security of the E91 protocol using arbitrary detectors can be reduced to that of the BB84 protocol using single photon detectors,
and that allows us to prove the asymptotic key generation rate $R=1-h((2+\sqrt2)p)-f_{\rm ec}h(p)$, with $p$ being the quantum error rate (QBER), $h(p)$ the binary entropy, and $f_{\rm ec}$ the efficiency of error correction.
This rate $R$ is higher than in the previous literature on the device-independent E91 protocol \cite{HRW10, MPA11, BCK12, RUV13, VV14,VV14Erata}, except the one assuming collective attacks, a very limited attack scenario \cite{PABGSV09} (see Figure \ref{fig:key_rate}).
For example, when the optimal error correcting code with $f_{\rm ec}=1$ is available, one can generate the secret key with the QBER up to 5.4\%.

For obtaining this result, we use the same physical assumptions as in Ref. \cite{MPA11}.
Namely, we assume that quantum mechanics is valid, and that Alice's and Bob's detectors are memoryless, i.e., different detectors operate on different Hilbert spaces.
In comparison with the other existing literature, these assumptions are weaker than in Ref. \cite{PABGSV09}, where collective attacks are assumed, but stronger than in Refs. \cite{BCK12,RUV13,VV14, VV14Erata}, where detectors are not necessarily memoryless.
They are also stronger than in Ref. \cite{HRW10}, which does not assume quantum mechanics.

Our security proof of the E91 protocol proceeds as follows.
In the first step, we convert the E91 protocol using arbitrary detectors into a simplified version where uncharacterized qubit detectors are used.
For this purpose we borrow the technique used in Ref. \cite{PABGSV09}, and the result is that, without loss of security, we may restrict ourselves to a protocol where Alice and Bob use qubit detectors, parameterized by complex numbers $\alpha,\beta$.
In the next step, we eliminate the $\alpha,\beta$-dependence by applying a bipartite squash operation $F_{\alpha,\beta}$, which is designed such that the CHSH measurement, jointly performed by Alice and Bob, is transformed to the phase error measurement of the BB84 type, also jointly performed by the two players.
$F_{\alpha,\beta}$ is also designed so that it leaves Alice's sifted-key measurement unchanged.
As a consequence, the original E91 protocol is transformed to the BB84 protocol, which can readily be shown secure by referring to the existing literature, e.g., \cite{SP00,RennerPhD,HT12,TLGR12}.

The crucial observation here is that the minimum entropy of Alice's sifted key depends only on the results of Alice's sifted-key measurement, and of the CHSH measurements on sample pulses.
No other measurements affect the sifted key as they are performed locally and remotely from it.
Hence for proving the security of the E91 protocol, it suffices to find a squash operation that properly transforms the CHSH and Alice's sifted-key measurement.
While the previous formulation based on the one-partite squash operation demands four conditions, corresponding to Alice's and Bob's choices of $x,z$ basis, which cannot be fulfilled in general, the bipartite generalization demands only two.
This is why this new setting realizes the security proofs that were not possible previously.

\section{Review of concepts regarding quantum key distribution and the security}

In this section, we clarify the notation and concepts to be used in this paper.
In particular, we explain the security criteria of QKD protocols, and review the previous method of the squash operation, restricted to one-partite measurement.

\subsection{$\varepsilon$-security and the smooth minimum entropy}
\label{sec:epsilon_security}

For the sake of simplicity, we restrict ourselves to entanglement-based QKD protocols.
Also for the sake of simplicity, we assume that the secret key length is constant; i.e., we only consider the type of protocols where Alice and Bob decide whether the protocol is aborted or not, by checking the measurement results of randomly chosen sample pulses, and when it continues, the generated secret key has a fixed bit length $l$.
Such protocols can typically be described as follows.

\begin{oframed}
\noindent {\bf Procotol 1 (PR1)}: Typical entanglement-based protocol using detector $M_c$.
\begin{enumerate}
\item (Quantum communication) Eve generates quantum state $\rho^{ABE}$, and sends its sub-states in ${\cal H}^A$ and ${\cal H}^B$, each consisting of $N$ tensor products of a Hilbert space ${\cal H}^M$, to Alice and Bob, respectively.
\item (Basis choice) For each pulse $i\in\{1,\dots,N\}$, Alice (resp., Bob) chooses bases $c_{A,i}\in\{z, x\}$ (resp., $c_{B,i}\in\{z, x\}$).
\item (Quantum measurement) Alice (resp., Bob) measures them using operators $M^A_{i}\left(r_{A,i}|c_{A,i}\right)$ (resp., $M^B_{i}\left(r_{B,i}|c_{B,i}\right)$), and record the results $r_{A,i}$ (resp., $r_{B,i}$).
\item (Determining whether continuing the protocol or not) Alice and Bob communicate through the public channel, and decides sample pulses $I_{\rm smp}\subset\{1,\dots,N\}$, and sifted key pulses $I_{\rm sif}\subset\{1,\dots,N\}$ in such a way that $I_{\rm smp}\cap I_{\rm sif}=\emptyset$.
By checking the measurement results of sample pulses $I_{\rm smp}$, they decide whether they continue or abort the protocol.

If they continue, Alice lets her measurement results of sifted key pulses $I_{\rm sif}$ be her sifted key $u$.
\item (Alice's privacy amplification)
Alice randomly selects hash function $f_{\rm pa}$ and announces it to Bob.
She then inputs her sifted key $u$ to $f_{\rm pa}$ and obtains her secret key $k=f_{\rm pa}(u)$ of $l$ bits, and stores it in ${\cal H}^K$.

\item (Bob's post-processing)
Alice calculates syndrome of her sifted key $u$, and announces it to Bob.
Bob lets his measurement results of sifted key pulses $I_{\rm sif}$ be his sifted key $u'$.
He corrects errors in $u'$ using syndrome, and by inputting the outcome to a hash function $f_{\rm pa}$, he obtains his secret key.
\end{enumerate}
\end{oframed}

In what follows, we denote all data announced in the public channel by a random variable $V$, Alice's secret key by $K$, and the final state corresponding to the initial state by $\rho^{ABE}$.
Eve eavesdrops information regarding the secret key $K$ by referring to $V$ and measuring her sub-state in ${\cal H}^E$.
The security against this attack is usually analyzed by defining the ideal state, and then evaluating how close it is with the actual state.
It is customary to define the ideal state to be where Alice's secret key $K$ seen from Eve is the perfectly uniform random source, i.e., $\rho_{\rm mix}^{K}\otimes \rho^{VE}$ with $\rho_{\rm mix}^K=2^{-l}\sum_{k=0}^{2^l-1}|k\rangle\langle k|$.
It is also customary to use the trace distance for evaluating the closeness with the actual state.
\begin{Dfn}
\label{def:security_criteria}
We say that a given QKD protocol is $\varepsilon$-secure if the following relation holds for an arbitrary attack by Eve:
\begin{equation}
d_1(\rho^{KVE}|VE):=\left\|\rho^{KVE}-\rho_{\rm mix}^{K}\otimes \rho^{VE}\right\|_1\le \varepsilon.
\label{eq:security_criteria}
\end{equation}
\end{Dfn}
As shown in Ref. \cite{BHLMO05}, this definition of security satisfies universal composability.

As emphasized by Renner \cite{RennerPhD}, in evaluating the trace distance $d_1(\rho^{KVE}|VE)$,
it is useful to consider the smooth minimum entropy $H_{\rm min}^{\varepsilon'}(\rho^{UVE}|VE)$ of sifted key $U$, because it allows the use of  mathematical tools similar to those of the Shannon theory.
This property is sufficient for bounding the trace distance from above, i.e.,
\begin{Lmm}
\label{lmm:leftoverhashing}
If function $f_{{\rm pa},V}$ for privacy amplification is randomly chosen from a universal$_2$ function family \cite{CW79}, then for any (sub-normalized) sifted key state $\rho^{UVE}$,
\begin{equation}
d_1(\rho^{KVE}|VE)=d_1(\rho^{f_{{\rm pa},V}(U)VE}|VE)
\le 2\varepsilon'+2^{-\frac12\left(H_{\rm min}^{\varepsilon'}(\rho^{UVE}|VE)-l\right)}.
\label{eq:quantum_leftover_hashing}
\end{equation}
\end{Lmm}
Here we denoted hash function $f_{\rm pa}$ by $f_{{\rm pa},V}$ in order to emphasize that it is determined uniquely by the public communication $V$.
We note that there is a useful generalization for this lemma using {\it dual} universal$_2$ functions \cite{FS08,TH13}, which allows the use of practically useful hash functions \cite{HT13}.

According to this lemma, once a lower bound on $H_{\rm min}^{\varepsilon'}(\rho^{UVE}|VE)$ is obtained for a given protocol, its security follows immediately.
For example,
if one can somehow prove that
\begin{equation}
H_{\rm min}^{\varepsilon/4}(\rho^{UVE}|VE)\ge l+2\log_2\frac1{\varepsilon}+6
\label{eq:security_intermsof_H_min}
\end{equation}
holds for an arbitrary attack by Eve, then Lemma \ref{lmm:leftoverhashing} guarantees that condition (\ref{eq:security_criteria}) of Definition \ref{def:security_criteria}, and thus the protocol is $\varepsilon$-secure.

As we restrict ourselves to entanglement-based protocols in this paper, once Eve fixes the initial state $\rho^{ABE}$,
the state $\rho^{UVE}$ describing Alice's sifted key and Eve is uniquely determined, as well as $\rho^{KVE}$ describing Alice's secret key and Eve.
This fact can be used to simplify the notation to some extent.
Define a (not necessarily trace preserving) completely positive map $\Pi_{\rm sif}$ for describing Alice's sifted key generation, $\Pi_{\rm pa}$ for her privacy amplification, and $\Pi_{\rm sec}=\Pi_{\rm pa}\circ\Pi_{\rm sif}$ for secret key generation.
Then $\rho^{UVE}$ can be denoted as $\rho^{UVE}=\Pi_{\rm sif}(\rho^{ABE})$, $\rho^{KVE}=\Pi_{\rm sec}(\rho^{ABE})$.
Here we use a convention that $\rho^{U,V=v',E}=\rho^{K,V=v',E}=0$ when protocol is aborted and no secret key is generated (i.e., $v'$ denotes a record of public communication that includes ``abort'').
We also use the notations, $\rho_*^{KVE}$, $\rho_*^{UVE}$, $\Pi_{{\rm sec},*}$, $\Pi_{{\rm sif},*}$, with symbol $*$ specifying the protocol or game used.
For example, $\rho_{\rm PR1}^{KVE}$ is the final state generated by Protocol 1 (PR1) from the initial state $\rho^{ABE}$, i.e., $\rho_{\rm PR1}^{KVE}=\Pi_{\rm sec,PR1}(\rho^{ABE})$.

In these notations, condition (\ref{eq:security_criteria}) of Definition \ref{def:security_criteria} can be rewritten as
\begin{equation}
\max_{\rho^{ABE}}d_1(\Pi_{\rm sec}(\rho^{ABE})|VE)\le\varepsilon.
\label{eq:security_criteria_rewritten}
\end{equation}
Similarly, eq. (\ref{eq:security_intermsof_H_min}), which is a sufficient condition for (\ref{eq:security_criteria_rewritten}), can be rewritten as
\begin{equation}
\min_{\rho^{ABE}}H_{\rm min}^{\varepsilon/4}(\Pi_{\rm sif}(\rho^{ABE})|VE)\ge l+2\log_2\frac1{\varepsilon}+6.
\end{equation}

\subsection{Previous method using squash operations}
\label{sec:one_partite_squash}

As mentioned in Introduction, the squash operation is a mathematical tool that translates the security of a given QKD system using  practical detectors into that of qubit-based protocol.
In this subsection we review this method, based on the results of Ref. \cite{TT08}.
For the sake of simplicity, we will continue to restrict ourselves to entanglement-based protocols, although the method presented below can also be applied to prepare-and-measure protocols. 

Consider a QKD system where Alice's and Bob's detectors are not necessarily ideal qubit detectors, and denote the Hilbert space of their input by ${\cal H}^{M}$.
For instance, for the threshold detector (see, e.g., Ref. \cite{TT08}), ${\cal H}^{M}$ is the Fock space representing multi-photons.
For the sake of simplicity, we will further assume that measurement basis $c$ is chosen from $\{x, z\}$, and that the measurement outcome is $r\in\{\pm1\}$.
We denote the corresponding POVM in ${\cal H}^M$ by $M(r|c)$.
For later convenience, we also define operator $M(c):=M(+1|c)-M(-1|c)$.
In this setting, the squash operation is defined as follows.
\begin{Dfn}[1-partite squash operation]
\label{dfn:one-partite_squash}
A squash operation is a quantum operation (i.e., a trace-preserving and completely positive (TPCP) map) $F:{\cal H}^M\to {\cal H}^2$,
satisfying,
\begin{eqnarray}
M(x)=F^\dagger(X),\label{eq:dfn:one-partite_squash_x}\\
M(z)=F^\dagger(Z),\label{eq:dfn:one-partite_squash_z}
\end{eqnarray}
where $X$, $Z$ denote the Pauli matrices of the $x,z$ basis.
\end{Dfn}
Here, $F^\dagger$ denotes the Hermitian conjugate of $F$, i.e., the operator satisfying ${\rm tr}(MF(\rho))={\rm tr}(F^\dagger(M)\rho)$ for arbitrary state $\rho$ and measurement $M$.

Definition \ref{dfn:one-partite_squash} demands that measuring any state with an arbitrary basis $c\in\{x,z\}$ using $M(c)=M(+1|c)-M(-1|c)$ is equivalent to performing squash operation $F$ on the state and then measuring the resulting qubit state with the Pauli operators $X,Z$.
If such operation $F$ exists, all measurements in Protocol 1 performed by Alice and Bob using $M(r|c)$ ($c\in\{X,Z\}$) can be decomposed into $F$ followed by the normal qubit measurements using $X,Z$.
In security proofs, there is no loss of generality in supposing that $F$ is conducted by the attacker, so the security of Protocol 1 above can be reduced to that of the following protocol \cite{TT08}.
\begin{oframed}
\noindent {\bf Protocol 2 (PR2)}: A qubit-based protocol.

Same as Protocol 1 except
\begin{enumerate}
\item Eve generates quantum state $\rho^{\bar{A}\bar{B}E}$, where ${\cal H}^{\bar{A}}$ and ${\cal H}^{\bar{B}}$ consists of qubit spaces.
Then she sends its sub-states in ${\cal H}^{\bar{A}}$ and ${\cal H}^{\bar{B}}$ to Alice and Bob, respectively.
\setcounter{enumi}{2}
\item Alice (Bob) measures pulse $i$ using the Pauli operators $X^A, Z^A$ (resp., $X^B, Z^B$) corresponding to basis $c_{A,i}=x,z$ ($c_{B,i}=x,z$), and record the results $r_{A,i}$ ($r_{B,i}$).
\end{enumerate}
\end{oframed}
In terms of the trace distance and the smooth minimum entropy, we have the following relations.
\begin{Lmm}
\label{lmm:one_partite_squash_operation}
Let $\rho^{KVE}_{\rm PR1}$ and $\rho^{KVE}_{\rm PR2}$ be the final states generated as a result of Protocol 1 and 2, respectively, then we have
\begin{equation}
\max_{\rho^{ABE}} d_1(\rho^{KVE}_{\rm PR1}|VE)\le \max_{\rho^{\bar{A}\bar{B}E}} d_1(\rho^{KVE}_{\rm PR2}|VE).
\end{equation}
Similarly, let $\rho^{UVE}_{\rm PR1}$ and $\rho^{UVE}_{\rm PR2}$ be the sifted-key states generated by Protocol 1 and 2, respectively, then
\begin{equation}
\min_{\rho^{ABE}}H_{\min}(\rho^{UVE}_{\rm PR1})\ge \min_{\rho^{\bar{A}\bar{B}E}}H_{\min}(\rho^{UVE}_{\rm PR2}).
\end{equation}
\end{Lmm}
In \ref{sec:proof_of_lemma2}, we give a formal proof of this lemma.

According to Lemma \ref{lmm:one_partite_squash_operation}, once the squash operation $F$ is shown to exist for a given detector $M(r|c)$, any security proof of a qubit-based protocol remains valid even when the qubit detectors are replaced with $M(r|c)$.
In other words, once $F$ is known to exist for a practical detector $M(r|c)$,  it always suffices to consider the simplified case where the ideal qubit detectors are used; all analyses related with $M(r|c)$ become unnecessary.
This is the advantage of considering the squash operation.

\section{Multi-partite squash operations}
\label{sec:multi-partite_squash_operations}

\subsection{Motivation}
As we have seen in the previous section, once the squash operation $F$ is shown to exist for a practical detector $M(r|c)$, it serves as a very useful tool for analyzing protocols involving $M(r|c)$.
At the present, however, $F$ is shown to exist for a relatively limited class of detectors, e.g., the threshold detector of the BB84 type measurement \cite{TT08,BML08,T10}, and of the six-state measurement with a passive basis choice \cite{BML08}, and a few others \cite{GBNAML14}.
In fact, even if we restrict ourselves to qubit measurements of the BB84 type, we can easily construct counterexamples of $F$:
\begin{Lmm}[limitation of the 1-partite squash operation for a qubit]
\label{lmm:no-go_theorem}
In the notation of Definition \ref{dfn:one-partite_squash}, let $M(x)=X_\alpha$, $M(z)=Z$ with $X_{\alpha}$ being the generalized $X$ operator, defined in \ref{app:Pauli_matrices}.
If $\alpha\ne i,-i$, there exists no squash operation $F$ satisfying conditions (\ref{eq:dfn:one-partite_squash_x}), (\ref{eq:dfn:one-partite_squash_z}).
\end{Lmm}
This can be verified readily by checking conditions (3a), (3b), (3c) of Ref. \cite{BML08}.
The lemma says that even if one uses perfectly sensitive qubit detectors $X_{\alpha},Z$, the corresponding $F$ does not exist unless the alignment of the $x, z$ axes (corresponding to $\alpha$) is also perfect\footnote{It should be noted that the lemma may not be true if the $x, z$ measurements are embedded in more than two dimensions, or if their sensitivities are not perfect. In fact this is why $F$ exists for many practical cases listed above the lemma.}.
Hence one cannot hope to apply the method of the 1-partite squash operation to general detectors and reduce the security proof to qubit spaces.
Note that this difficulty applies to any protocol using measurements of the BB84 type, including the E91 protocol.

In the rest of this paper, we show that this situation changes drastically by considering a generalized setting where multi-partite measurements are involved.
That is, while all previous studies of the squash operation were concerned with a detector used by a single player, we here consider a generalization including measurements performed jointly by two players or more, such as the Bell and the CHSH measurements.
This approach allows us to relax mathematical conditions required for the existence of the squash operation, such that they can be fulfilled for a wider class of measurements.

\subsection{Definition}
We write down the definition for the multi-partite case.
This is a simple generalization of the one-partite squash operation of the previous section.
Consider a situation where $n$ players $P_1,\dots,P_n$ agree on a basis choice $c$ and perform (possibly non-local) $n$-partite measurements in the Hibert state ${\cal H}^{P_1}\otimes\cdots\otimes {\cal H}^{P_n}$ using operator $M(r|c)$ to obtain an outcome $r$.
Here the basis choice $c$ can be a list $(c_1,\dots,c_n)$ consisting of $P_i$'s choices $c_i$, but is not limited to this type.
More generally, it may also specify non-local measurements, such as the CHSH measurement, denoted by $c={\rm CHSH}$.
We assume that $c$ is chosen from a predetermined set $C$.
We also assume there are measurement operators $m(r|c)$ defined for the same variables $r,c$, which operate in $n$-qubit spaces $\bar{\cal H}^{P_1}\otimes\cdots\otimes \bar{\cal H}^{P_n}$.
In this setting, the multi-partite squash operation is defined as follows.
\begin{Dfn}[$n$-partite squash operation]
The squash operation for $n$-partite measurements $M(r|c)$ and $m(r|c)$ is a quantum operation $F:{\cal H}^{P_1}\otimes\cdots\otimes {\cal H}^{P_n}\to \bar{\cal H}^{P_1}\otimes\cdots\otimes \bar{\cal H}^{P_n}$ which satisfies an equality
\begin{equation}
M(r|c)=F^\dagger(m(r|c))
\label{eq:multi_partite_squash_operation}
\end{equation}
or an inequality
\begin{equation}
M(r|c)\ge F^\dagger(m(r|c))
\label{eq:multi_partite_squash_operation_inequality}
\end{equation}
for each basis choice $c\in C$.
\end{Dfn}

In the following sections, we show that this generalized approach can be used to prove the security of the E91 protocol using any detectors.
This is an evidence that our approach indeed allows to apply the squash operation to a wider class of protocols or detectors than previously possible, including the counterexample given in Lemma \ref{lmm:no-go_theorem}.

\section{Application of bipartite squash operation: device-independent QKD protocol}

In order to demonstrate the effectiveness of the multi-partite squash operation, introduced in the previous section, we apply it to the device-independent security proof of the E91 protocol using arbitrary detectors, and obtain the device-independent key generation rate $R=1-h\left((2+\sqrt2)p\right)-f_{\rm ec}h(p)$, improved upon those obtained in the previous literature.

\subsection{Outline of the proof}

In this paper we only consider the types of post-processing algorithms which execute bit error correction and privacy amplification as independent processes and do not exploit their correlations.
Hence the security analysis amounts to evaluating the minimum entropy $H_{\min}(\rho^{UVE}|VE)$ associated with the joint state $U$ $\rho^{UVE}$ of Alice's sifted key and Eve's quantum state $VE$, conditioned on the result of the CHSH test performed by Alice and Bob.
Then by assuming that Alice's and Bob's detectors are memoryless, we can neglect back reactions from Bob's sifted key measurement (the $z'$-basis measurement), since it does not affect $\rho^{UVE}$, to be evaluated.
Hence it suffices to analyze how $\rho^{UVE}$ behaves under Alice's sifted-key measurement and the CHSH measurement, both consisting only of the $x$- and the $z$-basis measurements by Alice and Bob.

\subsubsection{Reduction to the qubit-based E91 protocol with uncharacterized $X$ measurement}
The first crucial step is to reduce this evaluation of $\rho^{UVE}$ to the case where only (generalized) qubit detectors are used.
Borrowing the argument of Ref. \cite{PABGSV09}, we can actually assume, without loss of generality, that input pulses to Alice and Bob are all qubits, and that they measure them using the generalized $x$, $z$ measurements in qubit space, defined by\footnote{Throughout the paper, we represent the Pauli matrices in the $y$-basis. For details see \ref{app:Pauli_matrices}.}
\begin{equation}
X_{\alpha}^A=
\left(
\begin{array}{cc}
0&\alpha\\
\alpha^*&0
\end{array}
\right),
\quad
X_{\beta}^B=
\left(
\begin{array}{cc}
0&\beta\\
\beta^*&0
\end{array}
\right),
\quad
Z^A=Z^B=
\left(
\begin{array}{cc}
0&1\\
1&0
\end{array}
\right),
\label{eq:reduced_detector_matrix_1}
\end{equation}
with $\alpha,\beta$ being complex numbers satisfying $|\alpha|=|\beta|=1$.
Parameters $\alpha$ and $\beta$ are chosen arbitrary by Eve and may vary for different qubit pairs $i$.
Accordingly, we may assume that Alice's sifted-key measurement and the CHSH measurement are
\begin{eqnarray}
M_{\alpha,\beta}(c_A=z)&=&Z^A\otimes \II^B,
\label{eq:qubit_Z_measurement_defined}\\
M_{\alpha,\beta}({\rm CHSH})&=&\frac14(Z^{A}\otimes Z^{B}+Z^{A}\otimes X^B_\beta
+X^A_\alpha\otimes Z^{B}-X^A_\alpha\otimes X^B_\beta).
\label{eq:def_M_alpha_beta_qubit}
\end{eqnarray}
The details of this reduction to qubit spaces are presented in Lemma \ref{lmm:decomposition_to_qubits} and in \ref{app:proof_reduction_to_qubit_measurement}.
In short, it suffices to evaluate $H_{\min}(\rho^{UVE}|VE)$, supposing that Alice and Bob jointly perform qubit-pair measurements (\ref{eq:qubit_Z_measurement_defined}), (\ref{eq:def_M_alpha_beta_qubit}), with $\alpha,\beta$ chosen arbitrarily by Eve.

\subsubsection{Bipartite squash operation}
The second step is to eliminate uncontrollable parameters $\alpha$, $\beta$, chosen by Eve, by a security reduction using the squash operation.
As we will prove in Theorem \ref{thm:squash_operation} of Section \ref{sec:bipartite_squash_operation}, there exists a bipartite squash operation $F_{\alpha,\beta}$ satisfying
\begin{eqnarray*}
F^{\dagger}_{\alpha,\beta}(Z^A\otimes \II^B)&=&Z^A\otimes \II^B,\\
F^{\dagger}_{\alpha,\beta}\left(\II+(\sqrt2-1)X^A\otimes X^B\right)&\ge&2M_{\alpha,\beta}({\rm CHSH}),
\end{eqnarray*}
where $X^A$ and $X^B$ are the usual Pauli operators: $X^A=X^A_{-i}, X^B=X^B_{-i}$, with $i=\sqrt{-1}$ being the imaginary unit.
By using this $F_{\alpha,\beta}$, the evaluation of $H_{\min}(\rho^{UVE}|VE)$ with qubit-pair measurement operators being (\ref{eq:qubit_Z_measurement_defined}), (\ref{eq:def_M_alpha_beta_qubit}) is reduced to the analysis of the BB84 protocol using sifted-key measurement $Z^A\otimes \II^B$ and the phase error measurement $X^A\otimes X^B$.

As a result, the security of the device-independent E91 protocol 
is reduced to that of the BB84 protocol, which has been fully analyzed in the existing literature, e.g., \cite{SP00,RennerPhD,HT12,TLGR12}.
These are the main ideas of our security proof.

\subsection{Description of the Ekert 1991 protocol}
We consider the following version of the E91 protocol.

\subsubsection{Assumptions}
We use two assumptions for the security proof in the subsequent sections.
The first assumption is that quantum mechanics is valid.
The second assumption is for detectors:
Recall that we only consider the type of protocol where Eve prepares the initial state first, and Alice and Bob measure it using $N$ detectors respectively, with $N$ denoting the number of raw key bits.
In this setting we assume that these $2N$ detectors are memoryless, or uncorrelated with each other.

The precise description of the second condition is as follows.
Let us use variable $P\in\{A,B\}$ to denote players Alice and Bob.
We assume that the Hilbert space representing player $P$'s incoming state is clustered as ${\cal H}^{P}={\cal H}^{P}_1\otimes\cdots\otimes {\cal H}^{P}_N$, and that detector $i\in\{1,\dots,N\}$ operates only in subspace ${\cal H}^{P}_i$.
In other words, we assume that the $i$-th detector of player $P$ takes the form
\begin{equation}
\II^P_{1}\otimes \cdots\otimes \II^P_{i-1}\otimes M^P_{i}\left(r|c\right)\otimes\II^P_{i+1}\otimes \cdots\otimes \II^P_{N},
\label{eq:condition_cluster}
\end{equation}
where $c$ denotes the basis choice, $r\in\{\pm 1\}$ the output, and $\II^P_{i}$ the identity operator of ${\cal H}^{P}_i$.
We emphasize here that $M^P_{i}$ with different $P$ or $i$ may be different from each other.
In what follows we consider the situation where this conditions is guaranteed somehow, e.g., by shielding or separating detectors from each other.

We also restrict ourselves to the case where each detector $M^P_{i}$ always outputs value $r\in\{\pm 1\}$, and there is no inconclusive events, i.e.,
\begin{equation}
M^P_{i}(+1|c)+M^P_{i}(-1|c)=\II_{i}^P\quad{\rm for}\quad\forall c.
\end{equation}
Note this is not a new physical assumption, since any $M^P_{i}$ can be transformed to this type, e.g., by making it a rule that player $P$ assigns a random number $\pm1$ to output $c$ when detector $i$ says inconclusive.

\subsubsection{Procedures of the protocol}

The actual E91 protocol proceeds as follows.

\begin{oframed}
\noindent {\bf E91 Protocol, or  Game 0 (G0):} The E91 protocol with memoryless uncharacterized detectors.
\begin{enumerate}
\item (Quantum communication) Eve generates $N:=\displaystyle{\frac{n}{1-\delta}{\left(\frac1{1-q}\right)^2}}$ photon pairs  and sends them to Alice and Bob, where $\delta>0$ and $q\le 1/2$.

\item (Basis choice)
\begin{enumerate}
\item (Choosing sample and sifted key pulse candidates) On each pulse $i\in\{1,\dots,N\}$, Alice labels $b_{A,i}\in\{{\rm smp},{\rm sif}\}$ (denoting sample or sifted key candidate)
with probabilities $1-q, q$. 
Bob also labels $b_{B,i}\in\{{\rm smp},{\rm sif}\}$ in the same manner.
\item (Alice's basis choices)
Alice does the following for each $i$: If $b_{A,i}={\rm sif}$, let the basis $c_{A,i}=z$.
If $b_{A,i}={\rm smp}$, let $c_{A,i}=x,z$ with probability 1/2.
\item (Bob's basis choice) Bob does the following for each $i$: If $b_{B,i}={\rm sif}$, let the basis $c_{B,i}=z'$.
If $b_{B,i}={\rm smp}$, let $c_{A,i}=x,z$ with probability 1/2.
\end{enumerate}

\item (Quantum measurement) Alice and Bob measure pulses using operators $M^A_{i}\left(c_{A,i}\right)$ and $M^B_{i}\left(c_{B,i}\right)$.
The results are recorded as $(r_{A,i},r_{B,i})\in \{\pm1\}^2$.

\item (Determining whether continuing the protocol or not)
\begin{enumerate}
\item Selection of sample and sifted key pulses:
\begin{enumerate}
\item Alice and Bob announce all their labels $b_{A,i},b_{B,i}$ and basis choices $c_{A,i},c_{B,i}$ ($i=1,\dots,N$).
\item Alice randomly selects $l_{\rm smp}:=\displaystyle{n\left(\frac{q}{1-q}\right)^2}$ pulses (resp., $n$ pulses) satisfying $b_{A,i}=b_{B,i}={\rm smp}$ (resp., $b_{A,i}=b_{B,i}={\rm sif}$), and registers those $i$ as sample pulses $I_{\rm smp}\subset\{1,\dots,N\}$ (resp., $I_{\rm sif}\subset\{1,\dots,N\}$).
If it is found that there are less than enough pulses satisfying the conditions, the protocol is aborted.
\end{enumerate}
\item Verification of the CHSH inequality:
\begin{enumerate}
\item Alice reveals $\{r_{A,i}\, |\, i\in I_{\rm smp}\}$ to Bob.
\item
Bob calculates
\begin{equation}
S=\frac1{l_{\rm smp}}\sum_{i\in I_{\rm smp}}(-1)^{r_{A,i}+r_{B,i}+t\left(c_{A,i},c_{B,i}\right)}
\label{eq:S_defined}
\end{equation}
where 
\begin{equation}
t(c_{A},c_{B}):=
\left\{
\begin{array}{lll}
1 &{\rm if}\ c_{A}=c_{B}=x,\\
0&{\rm otherwise}.
\end{array}
\right.
\label{eq:t_defined}
\end{equation}
\item If $S$ is less than a given predetermined threshold $S_0$, Bob announces that the protocol is aborted.
\end{enumerate}
\item Generation of the sifted key:

Alice lets her measurement results of sifted key pulses $I_{\rm sif}$ be sifted key $u$, and stores it in ${\cal H}^U$.
\end{enumerate}
\item (Alice's post-processing) 
Alice calculates syndrome $v_{\rm syn}$ of her sifted key for error correction, and announces it to Bob.
We assume that the syndrome length $|v_{\rm syn}|$ satisfy $|v_{\rm syn}|\le l_{\rm syn}$.

Alice also selects a universal hash function $f_{\rm cor}$ randomly with output length $|f_{\rm cor}(\cdot)|=\lceil \log(1/\varepsilon_{\rm cor})\rceil$, and announces $f_{\rm cor}$ along with the hash value $f_{\rm cor}(u)$ of sifted key $u$.

Alice selects another universal hash function $f_{\rm pa}$ with output length $l$ randomly, calculates secret key $k=f_{\rm pa}(u)$, and stores it in ${\cal H}^K$.

\item (Bob's post-processing) 

Bob measures his sifted key pulses in the $z'$ basis to obtain his sifted key, and obtains corrected key $u'$ by performing bit error correction using syndrome $v_{\rm syn}$.
Then he verifies its correctness by checking if the hash value $f_{\rm cor}(u')$ of Bob's sifted key equals $f_{\rm cor}(u)$ sent from Alice.
If they differ the protocol is aborted; otherwise he obtains secret key of $l$ bits by applying a privacy amplification on the sifted key.
\end{enumerate}
\end{oframed}

\subsubsection{Remarks on the protocol besides security}

In Steps 2 and 4(a), Alice and Bob choose sample and sifted key pulses randomly, and if they fail to assign enough numbers of pulses, the protocol is aborted.
This abortion due to pulse selections occurs probabilistically and independently of Eve's choice of the initial state $\rho^{ABE}$.
This probability can be bounded by using the Chernoff bound (see, e.g., \cite{Mitzenmacher}, Theorem 4.5) as
\begin{equation}
\Pr[{\rm abort\ in\ Step\ 4(a)}]\le 2e^{-(\delta q)^2/2}.
\end{equation}

Parameter $S$, calculated in Step 3 (b), corresponds to the average of outcomes of the CHSH measurements.
That is, according to constructions of Step 2 and 3, obtaining $S$ is equivalent to measuring each sample pulse $i\in I_{\rm smp}$ using
\begin{eqnarray*}
&&\frac14( M_i^{A}(z)\otimes M_i^{B}(z) +  M_i^{A}(z)\otimes M_i^{B}(x)\\
&&\quad + M_i^{A}(x)\otimes M_i^{B}(z) -  M_i^{A}(x)\otimes M_i^{B}(x))
\end{eqnarray*}
with the outcome $s_i$, and then calculating the average $S=(l_{\rm smp})^{-1}\sum_{i\in I_{\rm smp}}s_i$.
In what follows, we will often call $S$ the {\it CHSH parameter}.

The uses of hash value $f_{\rm cor}(u)$ in Steps 5 and 6 guarantees that this protocol is $\varepsilon_{\rm cor}$-correct (see, e.g., Ref. \cite{RennerPhD,TLGR12}).

\subsection{Security of the above protocol}

\begin{Thm}
\label{Thm:security_E91}
The E91 protocol above is $\varepsilon$-secure.
That is, let $\rho^{KVE}_{\rm E91}$ be the final state generated by the E91 Protocol (or Game 0) on the input of initial state $\rho^{ABE}$,
consisting of secret key $K$, public communication $V$, and Eve's sub-state in ${\cal H}^E$.
Then we have
\begin{equation}
d_1(\rho^{KVE}_{\rm E91}|VE)\le \varepsilon,
\end{equation}
when the secret key length $l$ is chosen to be
\begin{eqnarray}
l&=&n\left(1-h\left((1+\sqrt2)\left(\frac1{\sqrt2}-S_0\right)+\mu'\right)\right)\nonumber\\
&&\ \ -2l_{\rm smp}-l_{\rm syn}-\log_2\frac1{\varepsilon_{\rm cor}}-2\log_2\frac3{\varepsilon},
\end{eqnarray}
with $\delta S$ and $\mu$ defined by
\begin{eqnarray}
\mu'&=&\left(4\sqrt3(1+\sqrt2)+\sqrt{\frac{(n+l_{\rm smp})(l_{\rm smp}+1)}{nl_{\rm smp}}}\right)\sqrt{\frac{1}{l_{\rm smp}}\ln\frac6{\varepsilon}}\,.
\label{eq:parameter_mu_defined}
\end{eqnarray}
where $h(p)$ denotes the binary entropy: $h(p)=-p\log_2 p-(1-p)\log_2 (1-p)$.
\end{Thm}
This theorem can be obtained by using the leftover hashing lemma (Lemma \ref{lmm:leftoverhashing}), and letting $\varepsilon'=\varepsilon/3$ in the following lemma.
\begin{Lmm}
Let $\rho^{UVE}_{\rm E91}$ be the state generated as a result of Steps 1 through 4 of the E91 Protocol (or Game 0) on the input of initial state $\rho^{ABE}$, and $U$ be the random variable denoting the sifted key.
Then $\rho^{UVE}_{\rm E91}$ satisfies, for an arbitrary value of $0<\varepsilon'\le1$, we have
\label{lmm:H_min_E91}
\begin{eqnarray}
\lefteqn{H_{\rm min}^{\varepsilon'}(\rho^{UVE}_{\rm E91}|VE)}\nonumber\\
&\ge& n\left(1-h\left((1+\sqrt2)\left(\frac1{\sqrt2}-\left(S_0-\delta S\right)\right)+\mu\right)\right)\\
&&-2l_{\rm smp}-l_{\rm syn}-\log_2\frac1{\varepsilon_{\rm cor}},
\end{eqnarray}
where
\begin{eqnarray}
\delta S&=&\sqrt{\frac{48}{l_{\rm smp}} \ln \frac2{\varepsilon'}}\,,\\
\mu&=&\sqrt{\frac{n+l_{\rm smp}}{nl_{\rm smp}}\frac{l_{\rm smp}+1}{l_{\rm smp}}\ln\frac2{\varepsilon'}}\,.
\label{eq:parameter_mu_defined}
\end{eqnarray}
\end{Lmm}
The proof of this lemma is given in the next section.

\subsection{Key generation rate for the qubit-based implementation}
In the ideal implementation of Ekert 1991 protocol, the entanglement source always generates the Bell state
\begin{equation}
|\Psi^{+1}\rangle:=\frac1{\sqrt2}\left(|0_z\rangle^A|0_{z'}\rangle^B+|1_z\rangle^A|1_{z'}\rangle^B\right),
\end{equation}
which is then sent to Alice and Bob, and measured using (presumable) single photon detectors.
In this setting, it is customary to rotate Bob's $x,z$ bases by 45 degrees with respect to those of Alice's, such that $S$ attains its maximum value $2\sqrt2$ when channels are noiseless.
It is also customary to choose Bob's $z'$ basis to be aligned with Alice's $z$ basis, so that their sifted keys match in the noiseless case.
When channels are noisy, e.g, the depolarizing channels with error rate $p$, the average of $S$ is
\begin{equation}
S=\frac{1-2p}{\sqrt2}
\end{equation}
and the bit error rate $p_{\rm sif}$ of sifted key equals $p$.

\begin{Crl}
In the above setting using single photon detectors, the asymptotic key generation rate $R:=\lim_{n\to\infty}n/N$ satisfies
\begin{equation}
R
=1-h\left((2+\sqrt2)p\right)-f_{\rm ec}h(p),
\end{equation}
where $f_{\rm ec}$ is the efficiency of error correction, i.e., asymptotic syndrome length is $l_{\rm syn}=f_{\rm ec}h(p)$.
\end{Crl}
{\it Proof:} For example, by choosing probability $q$ of basis choices as $q=n^{-1/2+\epsilon}$, $\epsilon>0$, we have $\mu'\to 0$, $l_{\rm smp}/N\to 0$, $n/N\to 1$ for $n\to\infty$, and obtain the lemma.
\sq

\begin{figure}
\centering
\includegraphics[width=10.00cm, clip]{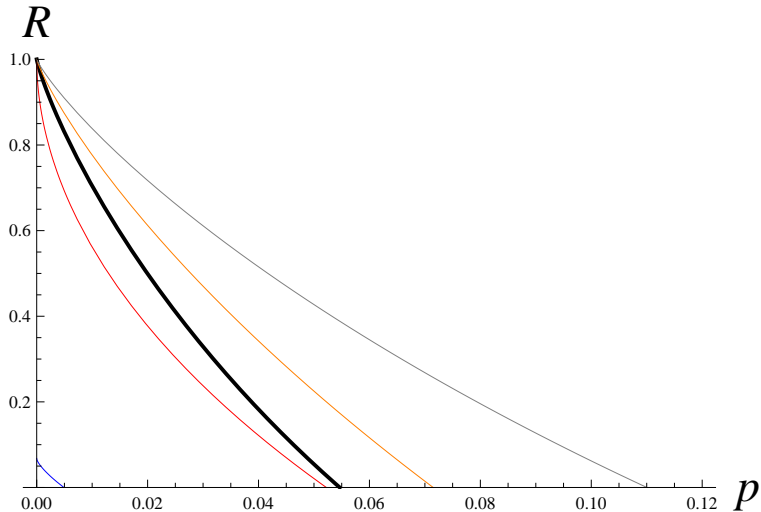}
\caption{(Color online) Asymptotic key generation rates $R$ versus quantum bit error rate $p$.
The bold black curve is our device-independent (DI) key generation rate for the E91 protocol using memoryless detectors.
The orange, red and blue curves are those obtained in Ref. \cite{PABGSV09} (DI against collective attacks), Ref. \cite{MPA11} (DI using memoryless detectors), and Refs. \cite{VV14,VV14Erata} (fully DI), respectively.
The gray curve is the rate of the usual device-{\it dependent} scenario using the Pauli measurements and one-way classical communications.
We let the bit error correction efficiency $f_{\rm ec}=1$ for all the cases.
Note that we achieve a higher rate throughout the domain than those under the same (red, \cite{MPA11}) or a severer (blue, \cite{VV14,VV14Erata}) scenarios.}
\label{fig:key_rate}
\end{figure}

Note that this rate $R$ is improved upon those in the previous literature on the device-independent E91 protocol \cite{HRW10, MPA11, BCK12, RUV13, VV14,VV14Erata}, except for the one assuming collective attacks \cite{PABGSV09} (see Figure \ref{fig:key_rate}).

\section{Proof of Lemma \ref{lmm:H_min_E91}}

The rest of this paper is devoted to the proof of Lemma \ref{lmm:H_min_E91}.
Our goal is to obtain a lower bound of $H_{\rm min}^\varepsilon(\rho^{UVE}_{\rm E91}|VE)$, with $\rho^{UVE}$ a sifted key state generated by Alice and Bob in Procotol 1, the E91 protocol.
As the direct analysis of such a practical system is usually cumbersome, we will use an indirect approach.
We convert the protocol to simpler procedures called {\it games}, which are defined as quantum operation which output a final state $\rho^{UVE}$ on the input of initial state $\rho^{ABE}$.
We use this terminology because some of the converted procedures can no longer be considered as a communication protocol; e.g., in Games $1,\dots,4$ below, a substitute player Charlie alone plays both of Alice's and Bob's parts, and there is no communication.

In the proof below, we start with the E91 protocol, also called Game 0, and repeat converting it to Games $1,2,\dots$, until we reach Game $n$, which is simple enough to analyze directly.
Games $i$ will also be abbreviated as G$i$ in what follows.
In order to be able to bound the minimum entropy $H_{\rm min}^{\varepsilon'}(\rho^{UVE}_{{\rm E91}}|VE)$ of Game $0$ by those of other games, we design the conversions such that the minimum entropy of  Game $i$ is not larger than that of the preceding game, Game $i-1$, possibly with a constant offset term $l_i\ge0$.

That is, we design conversions from Game $i-1$ to $i$ such that
\begin{equation}
\min_{\rho^{ABE}}H_{\rm min}^{\varepsilon'}(\rho^{UVE}_{{\rm G}i-1}|VE)\ge\min_{\rho^{ABE}} H_{\rm min}^{\varepsilon'}(\rho^{UVE}_{{\rm G}i}|VE)-l_{i},
\label{eq:trace_distance_between_transformed_protocols}
\end{equation}
is satisfied for all $1\le i\le n$.
Here $\rho^{UVE}_{{\rm G}i}$ denotes the sifted key state generated as the result of Game $i$, on the input of $\rho^{ABE}$, i.e., $\rho^{UVE}_{{\rm G}i}=\Pi_{{\rm G}i,{\rm sif}}(\rho^{ABE})$.
In this setting, it is immediate that the minimum entropy of the original protocol is bounded by that of Game $n$ as
\begin{eqnarray}
\min_{\rho^{ABE}}H_{\rm min}^{\varepsilon'}(\rho^{UVE}_{{\rm E91}}|VE)
&=&\min_{\rho^{ABE}}H_{\rm min}^{\varepsilon'}(\rho^{UVE}_{{\rm G}0}|VE)\nonumber\\
&\ge& \min_{\rho^{ABE}}H_{\rm min}^{\varepsilon'}(\rho^{UVE}_{{\rm G}n}|VE)-\sum_{i=1}^{n}l_i.
\label{eq:bound_on_protocol0}
\end{eqnarray}
Hence if a lower bound is obtained for the final Game $n$, then that of the E91 protocol follows automatically.
This type of situation is often described as `the security of the original protocol is {\it reduced} to that of Game $n$'.
We note that this approach using game transformations is not essentially new, and is implicit in the previous literature, such as \cite{RennerPhD,SP00}.

In our proof below, Game $0$ is the E91 protocol (Protocol 1), and we convert it to simpler Games $i$ ($i\ge1$) satisfying relation (\ref{eq:trace_distance_between_transformed_protocols}), until we reach Game 4, a security game of the BB84 protocol.

\subsection{Definition of the security game and the basic strategy of our proof}

As the first step, we define the following game.

\begin{oframed}
{\bf Game 1 (G1):} Security game of the E91 protocol.
\begin{enumerate}
\item (Input of the initial state)
Charlie receives from Eve an initial state $\rho^{ABE}\in{\cal H}^A\otimes{\cal H}^B\otimes {\cal H}^E$, with ${\cal H}^A$ and ${\cal H}^B$ each forming a space of $N':=n+l_{\rm smp}$ photon pulses.

\item (Selection of samples and measurement bases)
\begin{enumerate}
\item Charlie selects $l_{\rm smp}$ pulse pairs randomly, label them as sample pulses: $I_{\rm smp}\subset\{1,\dots,N'\}$, 
and announces it to Eve.
\item For all sample pulses $i\in I_{\rm smp}$, Charlie selects basis pairs $(c_{A,i},c_{B,i})\in\{(z,z),(z,x),(x,z),(x,x)\}$, each with probability $1/4$.
\end{enumerate}

\item (CHSH test)
Charlie measures sample pulses $i\in I_{\rm smp}$ using operator $M^A_{i}\left(c_{A,i}\right)\otimes M^B_{i}\left(c_{B,i}\right)$.
The results are recorded as $(r_{A,i},r_{B,i})\in \{\pm1\}^2$.
Then he calculates the CHSH parameter $S$ using (\ref{eq:S_defined}) and (\ref{eq:t_defined}), and if it is less than $S_0$, he announces to Eve that the protocol is aborted.

\item (Measurement of the sifted key)
Charlie measures each sifted key pulse $i\in I_{\rm sif}=\bar{I}_{\rm smp}=\{1,\dots,N'\}\setminus I_{\rm smp}$, using $M^A_{i}(z)\otimes \II^{B}$, and obtains a sifted key $U$ of $n$ bits.
Then he outputs the resulting state $\rho^{UVE}$.
\end{enumerate}
\end{oframed}

This game varies from Protocol 1 in four points:
\begin{itemize}
\item[i)] Alice's and Bob's procedures are all performed by a single substitute player, Charlie.
\item[ii)] For pulses that are neither samples nor sifted key, basis choices and measurement results are omitted.
\item[iii)] Charlie does not measure Bob's sifted key pulses.
\item[iv)] Charlie does not reveal basis choices of sample pulses, syndrome $v_{\rm syn}$ and hash value $f_{\rm cor}(u)$ to Eve, and keeps them secret.
\end{itemize}
It is straightforward to see that the first three modifications do not affect the output state $\rho^{UVE}$, nor the minimum entropy $H_{\min}^{\varepsilon'}(\rho^{UVE}|VE)$.
On the other hand, the fourth modification can affect $\rho^{UVE}$, since it erases some information available to Eve through public communication $V$, which is related with sifted key $U$ as well as basis choice of sample pulses.
In order to compensate the effect on $H_{\min}^{\varepsilon'}(\rho^{UVE}|VE)$ due to this lack of information properly,
we borrow results of Ref. \cite{RennerPhD} and prove the following lemma.

\begin{Lmm}
\label{lmm:Hmin_security_game}
For an arbitrary initial state $\rho^{ABE}$, we have
\begin{eqnarray}
H_{\rm min}^{\varepsilon'}(\rho^{UVE}_{{\rm E91}}|VE)
\ge H_{\rm min}^{\varepsilon'}(\rho^{UVE}_{{\rm G1}}|VE)-2l_{\rm smp}-l_{\rm syn}-\log_2\frac1{\varepsilon_{\rm cor}}.
\end{eqnarray}
\end{Lmm}
Note that this is an example of inequality (\ref{eq:trace_distance_between_transformed_protocols}).
The proof of this lemma is given in \ref{app:proof_min_entropy_compensation}.

\subsection{Reduction to a qubit-based game}

By borrowing the argument of Ref. \cite{PABGSV09}, we may assume, without loss of generality, that input pulses of Alice and Bob are all qubits.
\begin{Lmm}[c.f., Ref. \cite{PABGSV09}, Lemma 1]
\label{lmm:decomposition_to_qubits}
It is not restrictive to suppose that Eve sends to Alice and Bob a mixture $\rho_{AB}=\sum_{\alpha,\beta}p_{\alpha,\beta}\rho_{\alpha,\beta}$ of two-qubit states, together with a classical ancilla (known to her) that carries the values $\alpha,\beta$, and determines which measurements $M^A_{\alpha}(c_A)$ and $M^B_\beta(c_B)$ are to be used on $\rho_{\alpha,\beta}$.
\end{Lmm}
A proof sketch of this Lemma is given in \ref{app:proof_reduction_to_qubit_measurement}.
For the complete proof, we ask the reader to see Ref. \cite{PABGSV09}, Section 2.4.

This lemma states that, there is no loss of security (i.e., $H_{\min}(\rho^{UVE}|VE)$ does not increase), even if we restrict ourselves to the case where Eve generates a two-qubit state, accompanied by random variables $\alpha,\beta$, which are then measured by  Alice and Bob using operators $M^A_{\alpha}(c_A)$, $M^B_\beta(c_B)$.
With a suitable choice of bases, $M^A_{\alpha}(c_A)$, $M^B_\beta(c_B)$ can be expressed as $Z^A$, $Z^B$, $X^A_{\alpha}$, $X^B_{\beta}$ given in Eqs. (\ref{eq:reduced_detector_matrix_1}), (\ref{eq:reduced_detector_matrix_3}),
with $\alpha, \beta$ being complex numbers satisfying $|\alpha|=|\beta|=1$.
%
%
Thus the security of Game 1 can be reduced to that of the following game,
\begin{oframed}
{\bf Game 2 (G2):} (Qubit-based security game of the E91 protocol with uncharacterized detectors)

Same as Game 1 except
\begin{enumerate}
\item (Input of the initial state)
\begin{enumerate}
\item Eve selects complex numbers $\alpha_i$, $\beta_i$ ($|\alpha_i|=|\beta_i|=1$) for $i\in\{1,\dots,N'\}$, and announces them to Charlie.
\item Eve generates a state $\rho^{ABE}\in{\cal H}^A\otimes{\cal H}^B\otimes {\cal H}^E$, with each of ${\cal H}^A$ and ${\cal H}^B$ consisting of $N'$ qubit spaces, and gives it to Charlie.
\end{enumerate}

\setcounter{enumi}{2}
\item (CHSH test)
Charlie measures all $i\in I_{\rm smp}$ projectively using operators $Z^A\otimes Z^B$, $Z^A\otimes X^{B}_{\beta_i}$, $X^A_{\alpha_i}\otimes Z^B$, $X^A_{\alpha_i}\otimes X^B_{\beta_i}$ according to basis choices $(c^A_{i}, c^B_{i})=(z,z), (z,x), (x,z), (x,x)$.
The results are recorded as $(r^A_{i},r^B_{i})\in \{\pm1\}^2$.
Then he calculates the CHSH parameter $S$ using (\ref{eq:S_defined}) and (\ref{eq:t_defined}), and if it is less than $S_0$, he announces to Eve that the protocol is aborted.

\item (Measurement of the sifted key)
Charlie measures each sifted key pulse $i\in I_{\rm sif}=\bar{I}_{\rm smp}=\{1,\dots,N'\}\setminus I_{\rm smp}$ by using $Z^{A}\otimes \II^{B}$, and obtains a sifted key $U$ of $n$ bits.
Then he outputs the resulting state $\rho^{UVE}$.
\end{enumerate}
\end{oframed}
\noindent and we have the following lemma.
\begin{Lmm}
\label{lmm:Hmin_qubit_based_game}
For an arbitrary initial state 
\begin{equation}
\min_{\rho^{ABE}} H_{\rm min}^{\varepsilon'}(\rho^{UVE}_{{\rm G2}}|VE)\le \min_{\rho^{ABE}} H_{\rm min}^{\varepsilon'}(\rho^{UVE}_{{\rm G1}}|VE).
\end{equation}
\end{Lmm}

\subsection{Conversion to coherent CHSH measurements}

Unlike Alice and Bob in the actual E91 protocol, Charlie in Games 1 and 2 does not reveal basis choices of sample pulses to Eve.
Hence, performing projective measurements using operators $Z^A\otimes Z^B$, $Z^A\otimes X^B_{\beta}$, $X^A_{\alpha}\otimes Z^B$, $X^A_{\alpha}\otimes X^B_{\beta}$ in Step 4 in Game 2 (according to randomly chosen $c_{A},c_{B}\in\{x,z\}$) is equivalent to performing a measurement using a POVM
\begin{equation}
M_{\alpha,\beta}(\pm1|{\rm CHSH}):=\frac12\left(\II^A\otimes\II^B\pm M_{\alpha,\beta}({\rm CHSH})\right),
\label{eq:def_M_pm_measurement}
\end{equation}
where
\begin{eqnarray}
M_{\alpha,\beta}({\rm CHSH})&:=&\frac14(Z^{A}\otimes Z^{B}+Z^{A}\otimes X^B_{\beta}\\
&&+X^A_{\alpha}\otimes Z^{B}-X^A_{\alpha}\otimes X^B_{\beta}).\nonumber
\end{eqnarray}

It is straightforward to verify that $M_{\alpha,\beta}({\rm CHSH})$ can be diagonalized as
\begin{eqnarray}
M_{\alpha,\beta}({\rm CHSH})
=\sum_{a=\pm1}a\left(|\mu| |\Psi^a_\mu\rangle\langle \Psi^a_\mu|
+|\nu||\Phi^a_\nu\rangle\langle \Phi^a_\nu|\right)
\end{eqnarray}
where parameters $\mu,\nu$ are defined by
\begin{eqnarray}
\mu(\alpha,\beta)
&&=
\frac14(1+\alpha+\beta-\alpha\beta),
\label{eq:mu_defined}\\
\nu(\alpha,\beta)
&&=
\frac14(1+\alpha+\beta^*-\alpha\beta^*),
\label{eq:nu_defined}
\end{eqnarray}
and vectors $\{ |\Psi_\mu^{\pm1}\rangle, |\Phi_\nu^{\pm1}\rangle\}$ are the Bell states defined by
\begin{eqnarray}
|\Psi^{\pm1}_\mu\rangle&=&\frac1{\sqrt2}\left(|0_y\rangle^A|0_y\rangle^B\pm\frac{\mu}{|\mu|}|1_y\rangle^A|1_y\rangle^B\right),\\
|\Phi^{\pm1}_\nu\rangle&=&\frac1{\sqrt2}\left(|0_y\rangle^A|1_y\rangle^B\pm\frac{\nu}{|\nu|}|1_y\rangle^A|0_y\rangle^B\right).
\end{eqnarray}
Note here that parameters $|\mu|,|\nu|$ satisfy
\begin{equation}
|\mu|^2+|\nu|^2=\frac12.
\label{eq:mu_nu_relation}
\end{equation}
and thus $|\mu|, |\nu|\le\frac1{\sqrt2}$.

Note that POVM measurement $M_{\alpha,\beta}(\pm1|{\rm CHSH})$, defined by (\ref{eq:def_M_pm_measurement}), is equivalent to: a) performing projective measurement $M_{\alpha,\beta}({\rm CHSH})$ to obtain a result $\in \{\pm|\mu|, \pm|\nu|\}$, and then b) adding to it a noise factor $\in\{\pm1\pm|\mu|,\pm1\pm|\nu|\}$ chosen locally by Charlie by a certain probability distribution.
The projective measurement $M_{\alpha,\beta}({\rm CHSH})$ can further be decomposed as:
 a-1)  performing the Bell state measurement, 
 and then a-2) outputting $\pm|\mu|$ when $|\Psi^{\pm1}\rangle$ are measured, and $\pm|\nu|$ when $|\Phi^{\pm1}\rangle$.

Hence, as a result, Game 2 can be rewritten in the following equivalent form:
\begin{oframed}
{\bf Game 2' (G2'):} Same as Game 2 except
\begin{enumerate}

\setcounter{enumi}{2}
\item (CHSH test by POVM $M_{\alpha,\beta}(\pm1|{\rm CHSH})$)
Charlie performs the Bell measurement on each sample pulse $i\in I_{\rm smp}$ using the basis $\{ |\Psi^{\pm1}_{\mu_i}\rangle, |\Phi^{\pm1}_{\nu_i}\rangle\}$.
If he measures $|\Psi^a_{\mu_i}\rangle$, he outputs $s_{i}=b$ with probability $\frac12(1+ab|\mu_i|)$, where $a,b\in\{\pm1\}$,
and if he measures $|\Phi^a_{\nu_i}\rangle$, he outputs $s_{i}=b$ with probability $\frac12(1+ab|\nu_i|)$.

If the average $S=(l_{\rm smp})^{-1}\sum_{i\in I_{\rm smp}}s_{i}$ is below the threshold value $S_0$, he announces that the protocol is aborted.
\end{enumerate}
\end{oframed}

In this game the noise factors $\{\pm1\pm|\mu_i|,\pm1\pm|\nu_i|\}$ occur independently for each pulse $i$ with a probability independent of Eve's attack, so their perturbation to the average $S$ converge to zero rapidly with sample pulse number $l_{\rm smp}=|I_{\rm smp}|\to\infty$.
Hence the average $S$ can be replaced with the average of measurement results $\{\pm|\mu_i|,\pm|\nu_i|\}$ obtained by measurement $M_{\alpha_i,\beta_i}({\rm CHSH})$ up to a negligible error probability.
That is, for a new game defined as follows:
\begin{oframed}
{\bf Game 3 (G3):} Same as Game 2 except
\begin{enumerate}
\setcounter{enumi}{2}
\item (CHSH test by projective measurements with $M_{\alpha,\beta}({\rm CHSH})$)
Charlie measures each sample pulse $i\in I_{\rm smp}$ projectively with the basis $\{ |\Psi^{\pm1}_{\mu_i}\rangle, |\Phi^{\pm1}_{\nu_i}\rangle\}$, and outputs $s_{i}\in\{\pm|\mu_i|, \pm|\nu_i|\}$ respectively.
If the average $S=(l_{\rm smp})^{-1}\sum_{i\in I_{\rm smp}}s_{i}$ is below the threshold $S_0-\delta S$, he announces that the protocol is aborted.
\end{enumerate}
\end{oframed}
\noindent we have the following lemma:
\begin{Lmm}
\label{lmm:Hmin_G2_to_G3}
\begin{eqnarray}
\min_{\rho^{ABE}} H_{\rm min}^{\varepsilon'}(\rho^{UVE}_{{\rm G2}}|VE)
&\ge&\min_{\rho^{ABE}} H_{\rm min}^{\varepsilon'/2}(\rho^{UVE}_{{\rm G3}}|VE)\\
\frac{\varepsilon'}{2}&=&\exp(-l_{\rm smp}(\delta S)^2/48).
\end{eqnarray}
\end{Lmm}
The proof is straightforward, and is given in \ref{sec:locally_generated_randomness}.

\subsection{Bipartite squash operation}
\label{sec:bipartite_squash_operation}

The CHSH test of Game 3 still depends on parameters $\alpha$ and $\beta$, chosen arbitrarily by Eve.
Our next step is to eliminate this $\alpha,\beta$-dependence by converting them to a measurement consisting only of the usual Pauli operators $X, Z$.
For this purpose, we introduce a bipartite squash operation, $F_{\alpha,\beta}$, corresponding to $M_{\alpha,\beta}({\rm CHSH})$.

\begin{Thm}
\label{thm:squash_operation}
For any $\alpha$,$\beta$ satisfying (\ref{eq:mu_nu_relation}), there exists a squash operation $F_{\alpha,\beta}$ satisfying 
\begin{eqnarray}
F_{\alpha,\beta}^\dagger\left(Z^{A}\otimes \II^{B}\right)&=&Z^A\otimes \II^B.\label{eq:E91_squash_condition1}\\
F^{\dagger}_{\alpha,\beta}\left(\II+(\sqrt2-1)X^A\otimes X^B\right)&\ge&2M_{\alpha,\beta}({\rm CHSH}),
\label{eq:E91_squash_condition2}
\end{eqnarray}
\end{Thm}
The proof of this theorem is given in \ref{app:proof_theorem_bipartite_squash}.

Condition (\ref{eq:E91_squash_condition1}) guarantees that sifted key $U$ has the same probability distribution whether $F_{\alpha,\beta}$ is applied or not.
It is an equality between POVM elements, as used in the previous formalism of the 1-partite squash operation.
On the other hand, condition (\ref{eq:E91_squash_condition2}) deviates the previous formalism  in that it is a relation between observables, not POVMs, so we elaborate on it and justify its use below.

The CHSH test of Game 3 is equivalent to checking
\begin{equation}
\frac1{l_{\rm smp}}\sum_{i\in I_{\rm smp}}2M_{\alpha_i,\beta_i}({\rm CHSH})^{A_iB_i}\ge 2(S_0-\delta S),
\label{eq:Bell_test_Game3}
\end{equation}
Suppose now that Charlie performs the CHSH test using $F_{\alpha,\beta}^\dagger(\cdots)$ on the l.h.s. of (\ref{eq:E91_squash_condition2}), instead of $2M_{\alpha,\beta}({\rm CHSH})$ on the r.h.s., without changing the threshold $S_0-\delta S$.
Then the test condition (\ref{eq:Bell_test_Game3}) becomes changed to
\begin{equation}
\frac1{l_{\rm smp}}\sum_{i\in I_{\rm smp}}F^{\dagger}_{\alpha_i,\beta_i}\left(\II^{A_iB_i}+(\sqrt2-1)X^{A_i}\otimes X^{B_i}\right)\ge 2(S_0-\delta S).
\label{eq:Bell_test_Game4}
\end{equation}
It is easy to see that any input state $\rho^{USVE}$ that can pass the CHSH test (\ref{eq:Bell_test_Game3}) always passes the new test (\ref{eq:Bell_test_Game4}) because of (\ref{eq:E91_squash_condition2}).
In this case we say that the new test (\ref{eq:Bell_test_Game4}) is {\it weaker} than (\ref{eq:Bell_test_Game3}).

More generally, we say that the CHSH test using observable $M_1$ is stronger (weaker) than that using $M_2$, whenever $M_1\le M_2$ ($M_1\ge M_2$) and they share the same threshold value for $S$.
In this terminology, we can prove that a stronger CHSH test results in an equal or larger minimum entropy $H_{\min}(\rho^{UVE}|VE)$.
Recall that we assume memoryless detectors, and thus sifted key $U$ and sample bits $S$ belong to different vector spaces (different sectors of tensor product).
In this case the CHSH test works as a filter that transforms an input $\rho^{USVE}$ to the output $\rho^{UVE}$.
Hence if the same state $\rho^{USVE}$ is CHSH-tested using observables $M_1\le M_2$, the resulting states satisfy $\rho^{UVE}_1\le\rho^{UVE}_2$, which leads to $H_{\min}(\rho_1^{UVE}|VE)\ge H_{\min}(\rho_2^{UVE}|VE)$ due to the property of the minimum entropy.
A more rigorous argument is given in \ref{sec:improved_lemma}.

We advance our game transformation further by applying this argument to Game 3.
By replacing condition (\ref{eq:Bell_test_Game3}) of Game 3 with (\ref{eq:Bell_test_Game4}), we obtain the following game:\footnote{%
In Game 4, we divide $F^{\dagger}_{\alpha_i,\beta_i}\left(\II^{A_iB_i}+(\sqrt2-1)X^{A_i}\otimes X^{B_i}\right)$ of (\ref{eq:Bell_test_Game4}) into two steps: $F_{\alpha_i,\beta_i}$ in Step (i)(c) and the $X^{A_i}\otimes X^{B_i}$ measurement in Step (iii).
We made this division so that the transition from Game 4 to 5 in the next subsection becomes transparent.
Note that this division does not affect the argument of this subsection, since it preserves the relation between inputs $\rho^{ABE}$ and outputs $\rho^{UVE}$ of Game 4.}
\begin{oframed}
{\bf Game 4 (G4):} Same as Game 2 except
\begin{enumerate}
\item \label{game2doubleprimestep1}
(Input of the initial state)
\begin{enumerate}
\item Same as Game 2.
\item Same as Game 2.
\item Charlie applies squash operations $F_{\alpha_i,\beta_i}$ to qubit pairs $i\in\{1,\dots,N\}$ given by Eve
\end{enumerate}
\setcounter{enumi}{2}
\item (Phase error measurements by operator $X^A\otimes X^B$)
Charlie measures sample pulse $i\in I_{\rm smp}$ using operator $X^{A_i}\otimes X^{B_i}$ and obtains results $t_i=\pm1$.
He announces that the protocol is aborted whenever
\begin{equation}
\frac{1}{l_{\rm smp}}\sum_{i\in I_{\rm smp}}t_i
\ge \frac{S_0-\delta S-\frac12}{\frac1{\sqrt2}-\frac12}
\label{eq:game3''condition}
\end{equation}
\end{enumerate}
\end{oframed}
\noindent Due to (\ref{eq:E91_squash_condition2}), the CHSH test condition of this game, (\ref{eq:Bell_test_Game4}), is weaker than that of Game 3, (\ref{eq:Bell_test_Game3}).
Hence for the same input state $\rho^{USVE}$, the minimum entropies after the CHSH tests of each game satisfy $H_{\min}(\rho_{\rm G3}^{UVE}|VE)\ge H_{\min}(\rho_{\rm G4}^{UVE}|VE)$.
From this we obtain the following lemma:
\begin{Lmm}
\label{lmm:Hmin_G3_to_G4}
\begin{eqnarray}
\min_{\rho^{ABE}} H_{\rm min}^{\varepsilon'/2}(\rho^{UVE}_{{\rm G3}}|VE)
&\ge&\min_{\rho^{ABE}} H_{\rm min}^{\varepsilon'/2}(\rho^{UVE}_{{\rm G4}}|VE)
\end{eqnarray}
\end{Lmm}


Now the whole quantum operation performed by Charlie in Step (\ref{game2doubleprimestep1}) of Game 4 can be regarded as a bipartite squash operation.
Thus we can apply the same argument as in Section \ref{sec:one_partite_squash}, and reduce this game to the following one where there is no squash operation.
Also, by noting that the phase error rate of samples is given by $p=\frac12\left(1-{l_{\rm smp}}^{-1}\sum_{i\in I_{\rm smp}}t_i\right)$, we see that condition (\ref{eq:game3''condition}) is equivalent to $p\le (1+\sqrt2)\left(\frac1{\sqrt2}-\left(S_0-\delta S\right)\right)$.
Hence, for the following game,
\begin{oframed}
{\bf Game 5 (G5):} Same as Game 2 except
\begin{enumerate}

\item (Input of the initial state)
Eve generates a state $\rho^{ABE}\in{\cal H}^A\otimes{\cal H}^B\otimes {\cal H}^E$, with each of ${\cal H}^A$ and ${\cal H}^B$ consisting of $N'$ qubit spaces, and gives it to Charlie.

\setcounter{enumi}{2}
\item (Phase error measurement using $X\otimes X$) Charlie detects phase errors of sample pulses $i\in I_{\rm smp}$ using operator $X^A\otimes X^B$, and aborts the protocol whenever the phase error rate $p$ satisfies
\begin{equation}
p\le (1+\sqrt2)\left(\frac1{\sqrt2}-\left(S_0-\delta S\right)\right).
\end{equation}
\end{enumerate}
\end{oframed}
\noindent we obtain the following lemma.
\begin{Lmm}
\label{lmm:Hmin_bipartite_squash}
\begin{eqnarray}
\min_{\rho^{ABE}} H_{\rm min}^{\varepsilon'/2}(\rho^{UVE}_{{\rm G4}}|VE)
&\ge&\min_{\rho^{ABE}} H_{\rm min}^{\varepsilon'/2}(\rho^{UVE}_{{\rm G5}}|VE).
\end{eqnarray}
\end{Lmm}

\subsection{Calculation of the key generation rate}

Since Game 5 is equivalent to the BB84 protocol,
by applying the existing security proofs of the BB84 protocol \cite{SP00,RennerPhD,HT12,TLGR12}, we can bound $H_{\rm min}^{\varepsilon'/2}(\rho^{UVE}_{{\rm G5}}|E)$ from below.
For example, by using a simple formula derived in \cite{TLGR12} for the finite length case, we obtain the following lemma.
\begin{Lmm}
\label{lmm:key_rate_of_finite_sized_BB84}
The output of Game 5, $\rho^{UVE}_{\rm G5}$, satisfies
\begin{equation}
H_{\rm min}^{\varepsilon'/2}(\rho^{UVE}_{{\rm G4}}|VE)\ge n\left(1-h\left((1+\sqrt2)\left(\frac1{\sqrt2}-\left(S_0-\delta S\right)\right)+\mu\right)\right).
\end{equation}
\end{Lmm}

{\it Proof:}
We follow the gedankenexperiment approach used in the security proof of \cite{TLGR12}.
Suppose that Alice and Bob perform $x$ basis measurement on all the squashed qubits, and denote Alice's measurement result by $W$ and Bob's by $W'$.
Then their maximum entropy is bounded by the threshold $Q_{\rm tol}$ of phase error rate of sample bits as
\begin{eqnarray}
H_{\max}^{\varepsilon'/2}(W|W')&\le& nh(Q_{\rm tol}+\mu)\nonumber\\
&=&nh\left((1+\sqrt2)\left(\frac1{\sqrt2}-\left(S_0-\delta S\right)\right)+\mu\right)
\end{eqnarray}
with $\mu$ defined in (\ref{eq:parameter_mu_defined}).
Combining this inequality with the uncertainty relation for smooth entropies derived in \cite{TR11}
\begin{equation}
H_{\min}^{\varepsilon'/2}(U|VE)+H_{\max}^{\varepsilon'/2}(W|W')\ge n,
\end{equation}
we obtain the lemma.
Here we used the notation $H_{\min}^{\varepsilon'/2}(U|VE)=H_{\min}^{\varepsilon'/2}(\rho^{UVE}|VE)$.
\sq

\subsection{Proof of Lemma \ref{lmm:H_min_E91}}
By combining Lemmas \ref{lmm:Hmin_security_game}, \ref{lmm:Hmin_qubit_based_game}, \ref{lmm:Hmin_G2_to_G3} \ref{lmm:Hmin_G3_to_G4}, \ref{lmm:Hmin_bipartite_squash}, and \ref{lmm:key_rate_of_finite_sized_BB84}, we obtain Lemma \ref{lmm:H_min_E91}.

\section{Conclusion}

We proposed a generalization of the squash operations involving  multi-partite measurements, and demonstrated that it allows us to prove the security of a wider class of QKD systems than previously possible.
In particular, we applied our method to prove the device-independent security of the Ekert 1991 (E91) protocol, and improved the key generation rate.

Note that for the conventional formalism of the 1-partite squash operation, there are explicit counterexamples, such as the one given in Lemma \ref{lmm:no-go_theorem}.
Hence our result on the device-independent E91 protocol is a concrete evidence that our approach indeed allows one to apply the squash operation to a wider class of protocols or detectors than previously possible.

We do not yet know how wide it will eventually be. Neither do we see any explicit limitation.
In this sense, possible future directions would be to investigate whether the techniques developed here can be applied to the cases where detectors are not memoryless, and to where two way classical communications are used for post processing.
It is also interesting to reinterpret the existing security proofs of partially device-independent or device-dependent protocols (e.g., Refs. \cite{GM14,M13}) using our method, and to develop them further.

\ 

\noindent{\bf Acknowledgment}

The authors would like to thank Takuya Hirano for valuable comments.
This work was partially supported by the National Institute of Information and Communications Technology (NICT), Japan, and by ImPACT Program of Council for Science, Technology and Innovation (Cabinet Office, Government of Japan).

\appendix

\section{Notation of the Pauli matrices}
\label{app:Pauli_matrices}
Throughout the paper, we use the $y$-basis repersentation of the Pauli operators represented as
\begin{equation}
X=
\left(
\begin{array}{cc}
0&-i\\
i&0
\end{array}
\right),
\quad
Y=
\left(
\begin{array}{cc}
1&0\\
0&-1
\end{array}
\right),
\quad
Z=
\left(
\begin{array}{cc}
0&1\\
1&0
\end{array}
\right).
\end{equation}
These Pauli matrices are the same as those used in most papers and textbooks, except that they are represented in the $y$-basis: $|b_y\rangle = \frac1{\sqrt2}\left(|0_z\rangle+(-1)^bi|1_z\rangle\right)$, $b\in\{0,1\}$.
One can recover the usual forms by rewriting them in the $z$-basis $|b_z\rangle$.

We also introduce the generalized $X$ operator, parametrized by a complex number $\alpha$ satisfying $|\alpha|=1$, as 
\begin{eqnarray}
X_{\alpha}&=&
\left(
\begin{array}{cc}
0&\alpha\\
\alpha^*&0
\end{array}
\right).
\label{eq:reduced_detector_matrix_3}
\end{eqnarray}
Note that this interpolates between $X$ and $Z$: $X=X_{-i}$, $Z=X_{1}$ with $i=\sqrt{-1}$.

\section{Proof of Lemma \ref{lmm:one_partite_squash_operation}}
\label{sec:proof_of_lemma2}
Consider the following intermediate protocol:
\begin{oframed}
\noindent {\bf Protocol 1' (PR1')}: Intermediate protocol using squash operation.
This is same as Protocol 1 except
\begin{enumerate}
\item Eve generates quantum state $\rho^{ABE}$, and sends its sub-states in ${\cal H}^A$ and ${\cal H}^B$, each consisting of $N$ tensor products of ${\cal H}^M$, to Alice and Bob, respectively.
Alice and Bob then apply squash operation $F$ and stores the resulting qubit states in ${\cal H}^{\bar{A}}$ and ${\cal H}^{\bar{B}}$.
\setcounter{enumi}{2}
\item Same as Protocol 2.
\end{enumerate}
\end{oframed}
All operations in Protocol 1' are identical to Protocol 1.
The only difference is that the measurements $M_c$ performed in Step 3 of Protocol 1 are divided into two steps,
the squash operation $F$ of Step 1 and qubit measurements $\sigma_c$ of Step 3.
Hence we have the following relation:
\begin{equation}
\Pi_{\rm sif,PR1}\left(\rho^{ABE}\right)=\Pi_{\rm sif,PR1'}\left(\rho^{ABE}\right).
\label{eq:equivalence_protocol12}
\end{equation}

Next by definition, we have
\begin{eqnarray}
\Pi_{\rm sif, PR1'}(\rho^{ABE})&=&\Pi_{\rm sif, PR2}((F_{\rm tot}\otimes \II^E)(\rho^{ABE})),\\
F_{\rm tot}&=&(\bigotimes_iF_{A,i})\otimes(\bigotimes_jF_{B,j}).
\end{eqnarray}
That is, inputting $\rho^{ABE}$ to Protocol 1' is equivalent to applying squash operation $F_{\rm tot}\otimes \II^E$ on $\rho^{ABE}$ first and then inputting the resulting state to Protocol 2.
Hence we have
\begin{eqnarray}
\lefteqn{\min_{\rho^{ABE}}H_{\min}\left(\rho^{UVE}_{\rm PR1}|VE\right)}\nonumber\\
&=&\min_{\rho^{ABE}}H_{\min}\left(\Pi_{\rm sif,PR1}\left(\rho^{ABE}\right)|VE\right)\nonumber\\
&=&\min_{\rho^{ABE}}H_{\min}\left(\Pi_{\rm sif,PR1'}\left(\rho^{ABE}\right)|VE\right)\nonumber\\
&=& \min_{\rho^{ABE}}H_{\min}(\Pi_{\rm sif,PR2}((F_{\rm tot}\otimes \II^E)(\rho^{ABE})|VE)\nonumber\\
&=& \min_{\rho^{\bar{A}\bar{B}E}=(F_{\rm tot}\otimes \II^E)(\rho^{ABE})}H_{\min}(\Pi_{\rm sif,PR2}(\rho^{\bar{A}\bar{B}E})|VE)\nonumber\\
&\ge&\min_{\rho^{\bar{A}\bar{B}E}}H_{\min}(\Pi_{\rm sif,PR2}(\rho^{\bar{A}\bar{B}E})|VE)\\
&=&\min_{\rho^{ABE}}H_{\min}(\rho^{UVE}_{\rm PR2}|VE),
\label{eq:inequality_protocol23}
\end{eqnarray}
where the minimum on the fifth line is over all $\rho^{\bar{A}\bar{B}E}$ for which there exists $\rho^{ABE}$ satisfying $\rho^{\bar{A}\bar{B}E}=(F_{\rm tot}\otimes \II^E)(\rho^{ABE})$.

\section{Proof of Lemma \ref{lmm:Hmin_security_game}}
\label{app:proof_min_entropy_compensation}
We divide the random variable $V$ describing the public information available to Eve in the E91 protocol into two parts;
$V_1$ describing basis choice of sample bits, syndrome $v_{\rm syn}$ and hash value $f_{\rm cor}(u)$, and $V_2$ describing all the remaining part .
Then by using Eq. (3.21) of Ref. \cite{RennerPhD}, and noting that $V_1$ can be described in $2l_{\rm smp}+l_{\rm syn}+\log_2\frac1{\varepsilon_{\rm cor}}$ bits, we have
\begin{eqnarray}
\lefteqn{H_{\rm min}^{\varepsilon'}(\rho^{UVE}_{{\rm E91}}|VE)=H_{\rm min}^{\varepsilon'}(\rho^{UV_1V_2E}_{{\rm E91}}|V_1V_2E)}\nonumber\\
&\ge& H_{\rm min}^{\varepsilon'}(\rho^{UV_1V_2E}_{{\rm E91}}|V_2E)-H_{\max}(\rho^{V_2}_{{\rm E91}})\nonumber\\
&\ge& H_{\rm min}^{\varepsilon'}(\rho^{UV_1V_2E}_{{\rm E91}}|V_2E)-2l_{\rm smp}-l_{\rm syn}-\log_2\frac1{\varepsilon_{\rm cor}}.
\label{eq:Lemma4_1}
\end{eqnarray}
By noting that $F_2$ is a classical random variable, and by slightly modifying Lemma 3.1.9 of Ref. \cite{RennerPhD}, we have
\begin{eqnarray}
H_{\rm min}^{\varepsilon'}(\rho^{UV_1V_2E}_{{\rm E91}}|V_2E)\ge H_{\rm min}^{\varepsilon'}(\rho^{UV_2E}_{{\rm E91}}|V_2E)
=H_{\rm min}^{\varepsilon'}(\rho^{UVE}_{{\rm G1}}|VE).
\label{eq:Lemma4_2}
\end{eqnarray}
From (\ref{eq:Lemma4_1}) and (\ref{eq:Lemma4_2}), we obtain the lemma.
\sq

\section{Proof of Lemma \ref{lmm:decomposition_to_qubits}}
\label{app:proof_reduction_to_qubit_measurement}
We give a proof sketch of Lemma \ref{lmm:decomposition_to_qubits}.
For the complete proof, we ask the reader to see Section 2.4 of Ref. \cite{PABGSV09}.
Also keep in mind that we here only discuss Alice's detector, because the proof for Bob's detector can be given in exactly the same way.

First, we show that (Alice's) POVM measurements $M(r|c)$ ($c\in\{x,z\}$, $r\in\{\pm1\}$) can be rewritten as projective measurements in a Hilbert space, augmented by an ancilla (see, e.g., \cite{Nielsen-Chuang}).

\begin{Lmm}[Implicit in the proof of Lemma 1 \cite{PABGSV09}]
\label{lmm:projective_measurements}
Given POVM $M_{c,\pm1}:{\cal H}^M\to \{\pm1\}$ satisfying
\begin{equation}
M(+1|c)+M(-1|c)=\II^M\ \ {\rm for}\ c\in\{x,z\},
\label{eq:completenes_M_basis_xz}
\end{equation}
there exist an ancilla $\xi\in{\cal H}^{M'}$ and projection operators $P(r|c)$ in ${\cal H}^{M}\otimes{\cal H}^{M'}$ satisfying
\begin{equation}
M(r|c)(\rho)=P(r|c)(\rho\otimes \xi)
\end{equation}
for an arbitrary $\rho\in {\cal H}^M$, and
\begin{equation}
P(+1|c)+P(-1|c)=\II^{MM'}\ \ {\rm for}\ c\in\{x,z\}.
\label{eq:P_c_completeness}
\end{equation}
\end{Lmm}

{\it Proof:}
By using the completeness relation (\ref{eq:completenes_M_basis_xz}) for $M_c$, and the equivalence of POVM and general measurements (see, e.g., Section 2.2 of \cite{Nielsen-Chuang}),
we see that, for each basis $c\in\{x,z\}$, there exists an ancilla $\xi_c$ and projection operators $\tilde{P}(r|c)$ in ${\cal H}^{M}\otimes{\cal H}^{c}$, satisfying
$
M(r|c)(\rho)=\tilde{P}(r|c)(\rho\otimes \xi_c)
$
for an arbitrary state $\rho\in{\cal H}^M$.
Then by letting $\xi=\xi_x\otimes\xi_z$, $P(r|x)=\tilde{P}(r|x)\otimes \II^z$ and $P(r|z)=\tilde{P}(r|z)\otimes \II^x$, we obtain the lemma.
\sq

Hence Game 1 can be rewritten in the form where Charlie prepares ancillas $\xi$ for all detectors, and then measures the initial state $\rho^{ABE}$ together with $\xi$ using projections $P(r|c)$ satisfying (\ref{eq:P_c_completeness}).
We further modify this game such that $\xi$ is prepared by Eve, instead of Charlie, and call it Game 1'.
In this case, the value of minimum entropy $\min_{\rho^{ABE}} H_{\rm min}(\rho^{UVE}|VE)$, realized in this modified game, is never larger than in Game 1, since Eve has a larger choice of $\rho^{ABE}$.
Additionally, Game 1' can also be considered as a limited case of Game 1 where Charlie's measurements are $P(r|c)$.
Thus, as long as our goal is to bound $\min_{\rho^{ABE}} H_{\rm min}(\rho^{UVE}|VE)$ from below, there is no loss of generality in assuming that Charlie's detector are projections $P(r|c)$ satisfying (\ref{eq:P_c_completeness}).

By introducing operators $A_c:=P(+1|c)-P(-1|c)$ for $c=x,z$, this condition can be rewritten as ${A_x}^2={A_z}^2=\II^{MM'}$, for which the following lemma can be applied.
\begin{Lmm}[Ref. \cite{PABGSV09}, Lemma 2]
Let $A_x$ and $A_z$ be Hermitian operators with eigenvalues equal to $\pm1$ acting on a Hilbert space ${\cal H}$ of finite or countable infinite dimension.
Then we can decompose the Hilbert space ${\cal H}$ as a direct sum
\begin{equation}
{\cal H}=\bigoplus_{\alpha} {\cal H}_\alpha^2
\end{equation}
such that ${\rm dim} ({\cal H}_\alpha^2)\le2$ for all $\alpha$, and such that both $A_x$ and $A_z$ act within ${\cal H}_\alpha^2$, that is, if $|\psi\rangle\in {\cal H}_\alpha^2$, then $A_x|\psi\rangle\in {\cal H}_\alpha^2$ and $A_z|\psi\rangle\in {\cal H}_\alpha^2$.
\end{Lmm}
Hence operators $A_c$, as well as $P(r|c)$, can all be block diagonalized to two-qubit subspaces labeled by $\alpha$.
Hence $P(r|c)$ can be decomposed as a projective measurement that specifies subspace $\alpha$, followed by qubit measurements performed in ${\cal H}^\alpha$.
This means that the index $\alpha$ may be considered as a classical variable conveyed from Eve to the legitimate player.
This concludes the proof of Lemma \ref{lmm:decomposition_to_qubits}.

\section{Proof of Theorem \ref{thm:squash_operation}}
\label{app:proof_theorem_bipartite_squash}

Define an operator
\begin{equation}
M'_{\alpha,\beta}({\rm CHSH}):=M_{\alpha,\beta}({\rm CHSH})+2|\mu||\Psi_\mu^{-1}\rangle\langle\Psi_\mu^{-1}|+2|\nu||\Phi_\nu^{-1}\rangle\langle\Phi_\nu^{-1}|.
\end{equation}
It is straightforward to show that it satisfies
\begin{equation}
M'_{\alpha,\beta}({\rm CHSH})\ge M_{\alpha,\beta}({\rm CHSH}).
\label{eq:Mprime_ge_M}
\end{equation}
and that it can also be written as
\begin{eqnarray*}
M'_{\alpha,\beta}({\rm CHSH})
&=&\sum_{a=\pm1}\left(|\mu||\Psi_\mu^{a}\rangle\langle\Psi_\mu^{a}|+|\nu||\Phi_\nu^{a}\rangle\langle\Phi_\nu^{a}|\right)\\
&=&\frac{|\mu|+|\nu|}2\II^{AB}+\frac{|\mu|-|\nu|}2Y^A\otimes Y^B.
\end{eqnarray*}
As the coefficients satisfy $(|\mu|+|\nu|)^2+(|\mu|-|\nu|)^2=2(|\mu|^2+|\nu|^2)=1$ and $|\mu|+|\nu|\ge|\mu|-|\nu|$, the operator $M'_{\alpha,\beta}({\rm CHSH})$ can further be rewritten using an angular parameter $|\phi(\alpha,\beta)|\le\pi/4$ as
\begin{eqnarray}
M'_{\alpha,\beta}({\rm CHSH})
&=&\frac12\left(\cos\phi(\alpha,\beta)\II^{AB}+\sin\phi(\alpha,\beta)Y^A\otimes Y^B\right).
\end{eqnarray}

Then by noting that Inequality (\ref{eq:E91_squash_condition2}) is equivalent to
\begin{equation}
(1+\sqrt2)\left(\II^{AB}-2M_{\alpha,\beta}({\rm CHSH})\right)+F_{\alpha,\beta}^\dagger(X^A\otimes X^B)\ge0,
\end{equation}
and by using (\ref{eq:Mprime_ge_M}), we see that Inequality (\ref{eq:E91_squash_condition2}) holds if
\begin{eqnarray}
N_{\alpha,\beta}&:=&(1+\sqrt2)\left(1-2M'_{\alpha,\beta}({\rm CHSH})\right)+F_{\alpha,\beta}^\dagger(X\otimes X)\\
&=&(1+\sqrt2)\left((1-\cos\phi)\II^{AB}+\sin\phi Y^A\otimes Y^B\right)+F_{\alpha,\beta}^\dagger(X\otimes X)\ge0\nonumber.
\end{eqnarray}

Hence, in order to prove Theorem \ref{thm:squash_operation}, it suffices to construct $F_{\alpha,\beta}$ satisfying (\ref{eq:E91_squash_condition1}) and $N\ge0$ explicitly.
For that purpose, it suffices to construct $F_{\alpha,\beta}$ satisfying (\ref{eq:E91_squash_condition1}) and
\begin{eqnarray}
F_{\alpha,\beta}^\dagger(X^A\otimes X^B)&=&a Y^A\otimes Y^B,\label{eq:appendix_F_alpha_beta_XX}\\
a&=&-{\rm Sign}(\sin\phi)\times \min(1, (1+\sqrt2)|\sin\phi|),\label{eq:appendix_a_defined}
\end{eqnarray}
where ${\rm Sign}(\sin\phi)=\pm1$ according to the sign of $\sin\phi$.
Note that $|a|\le1$.

One can indeed construct $F_{\alpha,\beta}$ satisfying (\ref{eq:appendix_F_alpha_beta_XX}) and (\ref{eq:appendix_a_defined}), e.g., by i) applying 90 degree $Z$ rotation to both $A$ and $B$ so that  $X^A\otimes X^B\to Y^A\otimes Y^B$, and then ii) 180 degree $Z$ rotation to $B$ only so that $Y^A\otimes Y^B\to -Y^A\otimes Y^B$, with probability $(1-a)/2$.
Note that (\ref{eq:E91_squash_condition1}) holds automatically since only $Z$ rotations are used.
$N_{\alpha,\beta}\ge0$ can be verified as follows:
If $|\sin\phi|\le 1/(1+\sqrt2)$ then $N=(1+\sqrt2)(1-\cos\phi)\II^{AB}\ge0$.
On the other hand if $|\sin\phi|>1/(1+\sqrt2)$, we have
\begin{eqnarray}
N&=&(1+\sqrt2)(1-\cos\phi)\II^{AB}+((1+\sqrt2)|\sin\phi|-1) {\rm Sign}(\sin\phi)Y^A\otimes Y^B\nonumber\\
&\ge&(1+\sqrt2)(1-\cos\phi)-((1+\sqrt2)|\sin\phi|-1))\II^{AB}\nonumber\\
&=&(1+\sqrt2)(\sqrt2-\cos\phi-|\sin\phi|)\II^{AB}\ge0.
\end{eqnarray}

\section{Stronger Bell test does not decrease the smooth minimum entropy}
\label{sec:improved_lemma}
Here we will prove a lemma, whose meaning can be roughly stated as: If one replaces one's Bell test measurement with a stricter one, then the minimum entropy of the state after passing it becomes larger.
This may sound obvious intuitively, but we give a proof for the completeness of our presentation.

We begin by defining exactly what we want to prove.
Let $Q_i$ and $R_i$ be two Bell measurement operators, defined for each pulse pair $i\in I_{\rm smp}$;
$Q_i$ and $R_i$ may vary depending on $i$.
Assume that absolute values of their eigenvalues are bounded from above uniformly by a constant $M>0$ which does not depend on  $i$,
and that they satisfy
\begin{equation}
Q_i\le R_i\quad {\rm for\ any}\ i\in I_{\rm smp}.
\end{equation}
Define their averages to be $\hat{Q}:={l_{\rm smp}}^{-1}\sum_{i\in I_{\rm smp}}Q_i$, $\hat{P}:={l_{\rm smp}}^{-1}\sum_{i\in I_{\rm smp}}P_i$, then we also have
\begin{equation}
\hat{Q}\le\hat{R}.
\label{eq:hatQ_le_hatR}
\end{equation}
Hence the Bell test using $\hat{Q}$ is stricter than that using $\hat{R}$, when the same threshold $S$ is used.

Also consider a projection operator $P_{Q}$ that outputs 1 when the projective measurement $\hat{Q}$ outputs a value larger than or equal to $S$, and outputs 0 otherwise.
This operator, $P_{Q}$, works as a filter that erases the input state whenever the Bell measurement $\hat{Q}$ outputs a value smaller than $S$.
$P_{R}$ is also defined in the same way as the filter that erases the input whenever $\hat{R}$ outputs a value smaller than $S$.

With these operators, the state after the Bell test by operator $\hat{Q}$ is represented as
\begin{equation}
\rho^{UVE}_Q={\rm Tr}_{S}(\rho_Q^{USVE}),\label{eq:app_rhoUVE_Q_Tr_S_rho_QUSVE}
\end{equation}
where
\begin{equation}
\rho^{USVE}_Q=(\II^U\otimes P_Q\otimes \II^{VE})\rho^{USVE}(\II^U\otimes P_Q\otimes \II^{VE})^\dagger.\label{eq:app_rhoUSVE_defined}
\end{equation}
The state after the Bell test $\hat{R}$, i.e., $\rho^{UVE}_Q$, is also defined by replacing subscript $Q$ in (\ref{eq:app_rhoUVE_Q_Tr_S_rho_QUSVE}), (\ref{eq:app_rhoUSVE_defined}) with $R$.

In this setting, we want to prove that the minimum entropy of $\rho^{UVE}_Q$ (state after the stronger Bell test) is not smaller than that of  $\rho^{UVE}_Q$ (state after the weaker Bell test).
This can be stated exactly as follows:

\begin{Lmm}
Under the above setting, operators $P_Q$, $P_R$ commute with each other and satisfy
\begin{equation}
P_Q\le P_R.
\label{eq:P_Q_le_P_R}
\end{equation}
The smooth minimum entropy of the corresponding states $\rho^{UVE}_P$, $\rho^{UVE}_Q$ satisfy
\begin{equation}
H_{\min}^{\varepsilon}(\rho^{UVE}_Q|VE)\ge H_{\min}^{\varepsilon}(\rho^{UVE}_R|VE).
\label{eq:Hmin_rhoUVEQ_ge_Hmin_rhoUVER}
\end{equation}

\end{Lmm}

{\it Proof:}
Let $V_{Qb}$, $V_{Rb}$ $(b\in\{0,1\})$ be the eigenspace of $P_Q$, $P_R$ associated with eigenvalue $b$.
If there exists a state $|\psi\rangle\in V_{Q1}\cap V_{R0}$, $|\psi\rangle\ne0$, then we obtain an inequality $S\le \langle\psi|\hat{Q}|\psi\rangle\le\langle\psi|\hat{R}|\psi\rangle\le S-\delta S$,
with $S-\delta S$ being the largest of eigenvalues of $\hat{R}$ that are smaller than $S$.
But this is absurd so $V_{Q1}\cap V_{P0}=\{0\}$, which implies $[P_Q, 1-P_R]=0$ and thus $[P_Q, P_R]=0$.
Hence $P_Q, P_R$ are simultaneously diagonalizable, and (\ref{eq:P_Q_le_P_R}) follows immediately from (\ref{eq:hatQ_le_hatR}).

Since $[P_Q, P_R]=0$ and (\ref{eq:P_Q_le_P_R}), $P_Q=P_Q P_R$ holds, so
\begin{eqnarray}
\rho^{USVE}_Q&=&(\II^U\otimes P_Q\otimes \II^{VE})\rho_R^{USVE}(\II^U\otimes P_Q\otimes \II^{VE})^\dagger.
\end{eqnarray}
By noting that $P_Q$ is a completely positive map, we can apply the date processing inequality (\cite{TomamichelPhD}, Theorem 5.7) to $\rho_R^{USVE}$, and obtain (\ref{eq:Hmin_rhoUVEQ_ge_Hmin_rhoUVER}).
\sq

\section{Proof of Lemma \ref{lmm:Hmin_G2_to_G3}}
\label{sec:locally_generated_randomness}
We first prove the following lemma, and then use it to prove Lemma \ref{lmm:Hmin_G2_to_G3}.

\begin{Lmm}
\label{lmm:Azuma_Hoeffding}
Denote random variable $S$ in Game 2', Game 3 by $S_{\rm G2'}$, $S_{\rm G3}$ respectively.
Then
\begin{equation}
\Pr\left[\left|S_{\rm G2'}-S_{\rm G3}\right|\ge \delta S\right]\le\frac{\varepsilon''}2,
\end{equation}
where
\begin{equation}
\frac{\varepsilon''}2:=\exp(-l_{\rm smp}(\delta S)^2/48).
\end{equation}
\end{Lmm}

{\it Proof of Lemma \ref{lmm:Azuma_Hoeffding}:}
Denote random variable $s_i$ of Game 2', Game 3 by $s_{{\rm G2'},i}$, $s_{{\rm G3},i}$ respectively, then $S_n={l_{\rm smp}}^{-1}\sum_{i\in I_{\rm smp}}s_{{\rm G}n,i}$ for $n=2', 3$.
Also define a random variable $t_i:=s_{{\rm G2'},i}-s_{{\rm G3},i}$, then it follows that their expected value is zero: $\langle t_i\rangle=0$ for $\forall i\in I_{\rm smp}$.
One can also verify easily that $t_i,t_j$ of different pulse pairs $i,j$ are independent from each other, and that their differences satisfy $\left|t_i-t_{i-1}\right|\le2(1+\sqrt2)$.
Thus we can apply the Azuma-Hoeffding inequality (see, e.g., Theorem 12.4, \cite{Mitzenmacher}) to their average, $S_{\rm G2'}-S_{\rm G3}$, and obtain the lemma.
\sq

{\it Proof of Lemma \ref{lmm:Hmin_G2_to_G3}:}
By using Lemma \ref{lmm:Azuma_Hoeffding} and by noting that the Bell test measurements of Game 2' and in Game 3 commute with each other, we have
\begin{eqnarray}
\|\Delta\rho^{USVE}\|&\le& \varepsilon'' \label{eq:DeltarhoUSVE_le},\\
\rho_{\rm G2'}^{UVE}+\Delta \rho^{UVE} &\le& \rho_{\rm G3}^{UVE}.\label{eq:rho_G2UVE+Delta}
\end{eqnarray}
By using (\ref{eq:DeltarhoUSVE_le}) and by definition of smooth min-entropy, we have $H_{\min}^{\varepsilon'+\varepsilon''}(\rho_{\rm G2'}^{UVE}|VE)\ge H_{\min}^{\varepsilon'}(\rho_{\rm G2'}^{UVE}+\Delta \rho^{UVE}|VE)$.
By using (\ref{eq:rho_G2UVE+Delta}) and the data processing inequality (\cite{TomamichelPhD}, Theorem 5.7), we also obtain $H_{\min}^{\varepsilon'}(\rho_{\rm G2'}^{UVE}+\Delta \rho^{UVE}|VE)\ge H_{\min}^{\varepsilon'}(\rho_{\rm G3}^{USE}|VE)$.
Combining these two inequalities, we have $H_{\min}^{\varepsilon'+\varepsilon''}(\rho_{\rm G2'}^{UVE}|VE)\ge H_{\min}^{\varepsilon'}(\rho_{\rm G3}^{USE}|VE)$.
\sq

\ 

\bibliography{squash}

\end{document}